\documentclass[%
 reprint,
superscriptaddress,
 amsmath,amssymb,
 aps,
]{revtex4-2}

\usepackage{graphicx}
\usepackage{dcolumn}
\usepackage{bm}
\usepackage{hyperref}

\usepackage{array}
\usepackage{tabularx}

\begin{document}
\preprint{APS/123-QED}

\title{Deformation potential extraction and computationally efficient mobility calculations in silicon from first principles}

\author{Zhen Li}
\email{Zhen.Li.2@warwick.ac.uk}
 \affiliation{School of Engineering, University of Warwick, Coventry CV4 7AL, United Kingdom.}
 
\author{Patrizio Graziosi}%
 \affiliation{Institute of Nanostructured Materials, CNR, Bologna 40129, Italy.}
 
\author{Neophytos Neophytou}
 \affiliation{School of Engineering, University of Warwick, Coventry, CV4 7AL, UK.}

\date{\today}

\begin{abstract}

We present a first-principles framework to extract deformation potentials in silicon based on density-functional theory (DFT) and density-functional perturbation theory (DFPT). We compute the electronic band structures, phonon dispersion relations, and electron-phonon matrix elements to extract deformation potentials for acoustic and optical phonons for all possible processes. The matrix elements clearly show the separation between intra- and intervalley scattering in the conduction band, and quantify the strength of the scattering events in the degenerate bands of the valence band. We then use an advanced numerical Boltzmann transport equation (BTE) simulator that couples DFT electronic structures and energy/momentum-dependent scattering rates to compute the transport properties for electrons and holes. By incorporating ionized impurity scattering as well, we calculate the $n$-type and $p$-type mobility versus carrier density and make comparisons to experiments, indicating excellent agreement. The fact that the method we present uses well-established theoretical tools and requires the extraction of only a limited number of matrix elements, makes it generally computationally very attractive, especially for semiconductors with a large unit cell and lower symmetry.

\end{abstract}

\maketitle


\section{Introduction}

Over the last two decades, a myriad of new materials and their alloys were synthesized and characterized in the search for new and improved functionalities across application areas\cite{Stoller2008,De2013,Burschka2013,Basov2017,Wang2018,Davide2019,Gibertini2019}. For every material studied experimentally, there are a lot more unexplored possibilities that can provide high performance. This has over the last years triggered a theoretical effort into machine learning studies to identify the component \cite{Kim2018,Schmidt2018}, atomic structure \cite{Oliynyk2016,Ziletti2018}, and physical properties with optimal functionalities \cite{Rajan2018,Stanev2018,Yuan2018}. Furthermore, a significant advancement into $ab$ $initio$ methods and density-functional perturbation theory (DFPT) \cite{Togo2015,Giannozzi2020,Gonze2020}, enabled more accurate calculations without the use of empirical parameters. These are used to predict new materials through high-throughput screening or machine learning, explore reaction mechanisms, and provide understanding in experimental synthesis and characterization.

Electron-phonon (e-ph) scattering is a vital part of simulations for materials properties, and $ab$ $initio$ calculations are becoming a critical component, enabling the investigation of e-ph scattering processes. Traditionally, e-ph scattering is employed within transport methods such as the Boltzmann transport equation (BTE) \cite{Li2015,DSouza2020,Neophytou2020}, Monte Carlo \cite{Pop2004,Akturk2008}, Landauer-Buttiker method \cite{Sowa2018}, etc. In the case of electronic transport, the BTE can be easily solved with the constant relaxation time approximation (CRTA) \cite{Georg2018}. Further study of e-ph scattering can be carried out using analytical models \cite{Jacoboni1977,Lundstrom2000} based on deformation potential theory which is developed by Bardeen and Shockley \cite{Bardeen1950}. The deformation potential essentially describes the shift in the bands upon a change in the lattice caused by a perturbation from specific phonon modes, the ones that dominate the overall process. Recent theoretical methods and available software can now solve the BTE by including interactions from the entire phonon spectrum for materials beyond common semiconductors \cite{Ponce2016, Samsonidze2018,Deng2020}. Such approaches, however, are computationally extremely costly in DFPT, even with the acceleration of Wannier interpolations \cite{Giustino2007}, as they require a dense electronic and phononic mesh discretization \cite{Ponce2018}, which leads to a large number of possible combinations in the calculation of e-ph interaction. Although these works started in the 1980s \cite{Baroni1987,Gonze1992,Savrasov1992}, it is only recently expanded to complex materials as a result of the advancements in computational resources and software developments. 

Deformation potential theory is still instrumental for the calculation of low-field mobilities \cite{Yoder1993,Fischetti1996,Davide2019}, as it is computationally much more efficient. It is also used routinely in high-field calculations in semiconductor devices, still with adequate accuracy \cite{Fischetti1991,Fischetti19912}. It can provide an understanding of individual phonon processes, and it can be easily employed to derive analytical scattering rates. These can then be used, for example, within device transport simulators \cite{Belarbi2016,Wu2018}, and in general when e-ph scattering needs to be combined with other scattering mechanisms, such as for nanostructured materials \cite{Fu2017}, or highly doped materials and alloys for which ionized impurity scattering \cite{Graziosi2019} and alloy scattering are important. Such methods are routinely employed for transistor devices and thermoelectric materials \cite{Graziosi2020}. The use of deformation potentials can allow for the flexibility and computational robustness that these simulators require.

In this paper, we use first-principles calculations based on DFT and DFPT to extract the deformation potentials for one of the common semiconductors, silicon (Si). Deformation potentials for Si have been used for decades now and are extracted from less advanced calculations or experimental measurements. Here, however, we perform DFPT to verify these numbers and obtain a deeper insight into the e-ph scattering processes in Si. The deformation potentials are derived from the e-ph interactions by considering the coupling of electrons/holes in an initial state to those of certain final scattering states under the perturbation induced by the dominant acoustic and optical phonon modes. We explore both $n$- and $p$-type carrier types. The method we present can be generally applied to other solid-state semiconductors and insulators, beyond the common Si material we focus on in this paper, by using the average optical phonon mode energy \cite{Samsonidze2018} and average optical deformation potential for a general semiconductor with multiple optical phonon branches. Using the extracted deformation potentials, we then extract the mobility of Si, with excellent agreement to experiment. We emphasize here that the use of Si in this work plays the role of a benchmark material to validate and establish the method we describe, since we can compare against numerous available data. However, the intention of the paper is to go beyond Si, and validate a generic and highly efficient computational method for mobility calculations. 

The paper is presented as follows: In Sec. II, we provide a description of the theoretical background for the extraction of deformation potentials. In Sec. III, we present the matrix elements and deformation potential extraction for Si. In Sec. IV, we calculate the scattering rate and mobility for holes and electrons using the extracted deformation potentials. In Sec. V, we discuss the results. Finally, in Sec. VI, we conclude.  

\section{Theory and Methods}

The mobility of semiconductors is determined by the e-ph scattering processes. The main processes commonly considered in theory and simulations are the scattering of electrons/holes with acoustic, optical, and polar optical phonons, when applicable. Phonons perturb the lattice, resulting in changes in the band structure of the semiconductor through perturbations in the crystal potential, which determines electronic transport. Generally, there are two main types of interactions between electrons and phonons, i.e., the deformation potential interaction, which describes the relation between the atoms displacement and the potential change near the displaced atoms, and the Fr\"{o}hlich interaction \cite{Frohlich1954}, which is related to the long-range electric fields in polar materials. Here we will focus on the deformation potential interaction, active in all solids (and dominant in Si).

In the original deformation potential theory by Bardeen and Shockley \cite{Bardeen1950}, the long-wavelength acoustic phonons (whose wavevector ${\bf{q}}$ $\rightarrow$ 0) are assumed to dominate the e-ph scattering mechanisms of electrons and holes in non-polar semiconductors. In this case, the long-wavelength acoustic phonons generate the atomic displacements and volume dilatation in the crystal, which shifts the electronic band energies. The acoustic deformation potential (ADP), which describes the relation between the energy shift and the volume expansion coefficient, can be computed by:
\begin{eqnarray}
D_{{\rm{ADP}},n\bf{k}}=V\frac{\partial E_{n\bf{k}}}{\partial V},
\label{eq1}
\end{eqnarray}
where $V$ is the volume of the unit cell, and $E_{nk}$ is the electronic eigenvalue for a band with index $n$ and wavevector $\bf{k}$ at the valance band maximum (VBM) or conduction band minimum (CBM). The qualitative meaning is that a high carrier mobility results from a small band shift with dilatation. However, the deformation potential $D_{\rm{ADP}}$ computed in this way, relying on a semi-empirical approach, does not consistently reproduce mobility measurements, and in general, lacks predictive power \cite{Giustino2017}. For quantitative prediction of the deformation potentials, which also takes into account more complex scenarios such as the effects of shear deformations and optical phonons, an $ab$ $initio$ self-consistent DFT calculation method \cite{Runge1984,Car1985} is needed to describe, both, lattice dynamics and electronic band structures. This also has the advantage of accounting for the screening of the ionic potential by the valence electrons automatically \cite{Vandenberghe2015}.

Here we use DFT and DFPT calculations to obtain the electronic band structures, phonon dispersion relations, and e-ph matrix elements entirely from the first principles. The key item is the determination of the e-ph matrix elements, which measure the coupling strength of the e-ph interactions. Specifically, the matrix element $g_{mn}^v (\bf{k},\bf{q})$ is the electronic response associated with a transition process where a Bloch electron at a state with band index $n$ and wavevector $\bf{k}$ scatters into a new state with band index $m$ and wavevector $\bf{k+q}$. This is facilitated by an atomic perturbation as a result of a phonon with mode index $\nu$ and crystal momentum $\bf{q}$. The matrix element can be determined using the variation formulation in DFPT as \cite{Savrasov1994, Liu1996}
\begin{eqnarray}
g_{mn}^\nu ({\bf{k},\bf{q}})=\sqrt {\frac{\hbar}{2m_0\omega_{\nu\bf{q}}}} M_{mn}^\nu ({\bf{k},\bf{q}}),
\label{eq2}
\end{eqnarray}
where $m_0$ is the sum of the masses of all the atoms in the unitcell, $\omega_{\nu\bf{q}}$ is the specific phonon frequency, and $M_{mn}^\nu (\bf{k},\bf{q})$ is defined as
\begin{eqnarray}
M_{mn}^\nu ({\bf{k},\bf{q}}) =\langle {\psi_{m{\bf{k+q}}}}({\bf{r}})|{\delta _{\nu {\bf{q}}}}V({\bf{r}})|{\psi _{n{\bf{q}}}}({\bf{r}}) \rangle,
\label{eq3}
\end{eqnarray}
where $\psi _{m{\bf{k+q}}}({\bf{r}})$ and $\psi_{n{\bf{q}}}( {\bf{r}} )$ are the electronic wavefunctions for band $m$ with wavevector $\bf{k+q}$ and band $n$ with wavevector $\bf{k}$, respectively, which are extracted from DFT calculations. The perturbing potential ${\delta _{\nu{\bf{q}}}}V( {\bf{r}})$ is associated with the phonon of branch index $\nu$, wavevector $\bf{q}$, and frequency $\omega_{\nu\bf{q}}$, which can be computed by DFPT. The details of how to extract the matrix element $g_{mn}^\nu (\bf{k},\bf{q})$ from DFPT can be found in Appendix A. Based on the e-ph matrix element for individual transitions, we derive below the deformation potential for acoustic and optical phonons.

\subsection{Acoustic deformation potential}
\begin{figure}[tbp]
\includegraphics[width=0.4\textwidth]{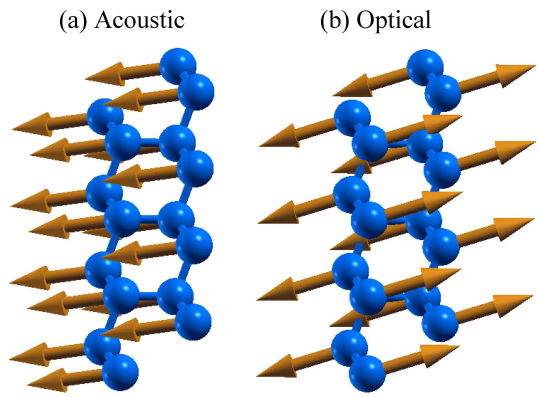}
\caption{Schematics of (a) acoustic and (b) optical vibration modes in the long-wavelength limit for Si visualized by forces on atoms.}
\label{fig1}
\end{figure}

The band structure is determined by the crystal potential, which is influenced by changes in the lattice spacing. The acoustic phonons in the long wave-length limit displace neighboring atoms in the same direction [see Fig. \ref{fig1}(a)], and thus the change of the lattice spacing is produced by the strain $\nabla \cdot \bf{u}$, where $\bf{u}$ is the displacement of atoms. In this case, the perturbing potential from acoustic phonons is approximately proportional to the strain as \cite{Lundstrom2000}
\begin{eqnarray}
V_{\rm{e\text{-}ph}}  = D_{\rm{ADP}} \nabla \cdot \bf{u},
\label{eq4}
\end{eqnarray}
where the proportionality constant $D_{\rm{ADP}}$ (in units of energy, eV) is the acoustic deformation potential (considered in general to be a constant \cite{Bardeen1950}). The system Hamiltonian can be decomposed as the original system’s Hamiltonian and a term related to the change in the system due to displacements of the nuclei by small amounts from their equilibrium positions. Within the harmonic approximation, the atomic displacement vectors, which diagonalize the altered component of the Hamiltonian of the crystal, can be expressed in terms of plane waves similar to the Bloch functions for electrons in a crystal \cite{Peter2010}. The $u_{\bf{q}}$, which is the displacement from equilibrium of an ion in a unit cell specified by the lattice vector $\bf{R}$, is related to the displacement $A_{\bf q}$ of a corresponding ion in the unit cell located at the origin by a Bloch wave of the form \cite{Satta1989,Peter2010}
\begin{eqnarray}
u_{\bf{q}} ({\bf{R}},t)=A_{\bf{q}} e^{+ i({\bf{q}} \cdot {\bf{R}} -\omega t)} + A_{\bf{q}} e^{-i({\bf{q}} \cdot {\bf{R}} -\omega t)},
\label{eq5}
\end{eqnarray}
where ${\bf{q}}$ and $\omega$ are the phonon wave vector and frequency, respectively. Here, for acoustic phonons, the strain $\nabla \cdot u_{\bf{q}} ({\bf{R}},t)$ can be written as $|{\bf{q}}| u_{\bf{q}} ({\bf{R}},t)$ in the case of small $u$. The $\delta_{\nu \rm{q}} V(r)$ term in $M_{mn}^\nu (\bf{k},\bf{q})$ in Eq. (\ref{eq3}) is proportional to the derivatives of the Kohn-Sham potential $V(r)$ with respect to changes in the atomic positions. Considering that the wavefunction overlap $\langle {\psi_{m{\bf{k+q}}}}({\bf{r}})|{\psi _{n{\bf{q}}}}({\bf{r}}) \rangle$ can be regarded as 1 at small $\bf{q}$ \cite{Lundstrom2000,Murphy2018}, the perturbing potential can be expressed as $V_{\rm{e\text{-}ph}} = M_{mn}^\nu ({\bf{k},\bf{q}})\cdot u_{\bf{q}}({\bf{R}},t)$. Thus, the acoustic deformation potential $D_{\rm ADP}$ can be calculated from Eq. (\ref{eq4}) as
\begin{eqnarray}
D_{\rm{ADP}} =\frac{M_{mn}^\nu ({\bf{k},\bf{q}})\cdot u_{\bf{q}}({\bf{R}},t)}{|{\bf{q}}|u_{\bf{q}}({\bf{R}},t)} =\frac{M_{mn}^\nu ({\bf{k},\bf{q}})}{|{\bf{q}}|}.
\label{eq6}
\end{eqnarray}

In the limit of small $\bf{q}$ (long wavelength phonons), this quantity $D_{\rm{ADP}}$ is the slope of $M_{mn}^\nu (\bf{k},\bf{q})$ with respect to the $|\bf{q}|$ of phonon eigenvector, and is refereed to as the first-order deformation potential. In order to obtain this from calculations, we consider an initial electronic state $\bf{k}$ (e.g., the top of the valence bands), and phonons (e.g., on the LA branch) along a high-symmetry direction (e.g., $\Gamma$-X). For different values of ${\bf{q}}$ along that phonon branch we compute the $M_{mn}^\nu (\bf{k},\bf{q})$ between the initial electronic state at $\bf{k}$, the final state at $\bf{k'= k+q}$ and the phonon at $\bf{q}$. Performing this calculation for different values of $\bf{q}$, allows us to obtain the deformation potential from the corresponding slope of the approximately straight line formed. Once the acoustic deformation potential is extracted, the scattering rate $|S_{\bf{k,k'}}^{\rm{ADP}}|$, which is the transition rate between the initial $\bf{k}$ and final $\bf{k'}$ states, can be computed using Fermi’s golden rule as \cite{Lundstrom2000,Neophytou2020}
\begin{eqnarray}
|S_{\bf{k,k'}}^{\rm{ADP}}|=\frac{\pi}{\hbar} D_{\rm{ADP}}^2 \frac{k_B T}{\rho v_s^2} \textsl{g}_{\bf{k'}},
\label{eq7}
\end{eqnarray}
where $\rho$ is the mass density, $v_s$ is the velocity of phonons in the branch used, and $\textsl{g}_{\bf{k'}}$ is the density of states at the final state. It is common in semiconductor electronic transport treatment to use a single deformation potential value for all the states in a specific band for simplicity, which is well justified as we will discuss below \cite{Lundstrom2000, Jacoboni1977}. In that case, $v_s$ becomes the sound velocity of the material.  

\subsection{Optical deformation potential}
The deformation potential theory can be extended to optical phonons, which arise when there are two or more atoms in the unit cell \cite{Harrison1956}. The neighboring atoms are displaced in opposite directions [see Fig. \ref{fig1}(b)], introducing a change in lattice spacing. In this case, it is the distances between the basis atoms, which disturb the surrounding lattice potential, acting as a scattering source on the electrons. Therefore, the perturbing potential is proportional to the atomic displacement as \cite{Potz1981}
\begin{eqnarray}
V_{\rm{e\text{-}ph}}  = D_{\rm{ODP}} \bf{u},
\label{eq8}
\end{eqnarray}
where $D_{\rm ODP}$ (in units of energy per unit length, eV/m) is the optical deformation potential (ODP). Compared to $V_{\rm{e\text{-}ph}} = M_{mn}^\nu ({\bf{k},\bf{q}})\cdot u_{\bf{q}}({\bf{R}},t)$, we can find that the $M_{mn}^\nu (\bf{k},\bf{q})$ is directly the $D_{\rm{ODP}}$. Thus, $D_{\rm ODP}$ is referred to as the zero-order deformation potential, which can be calculated as
\begin{eqnarray}
D_{\rm{ODP}}  = M_{mn}^\nu ({\bf{k},\bf{q}}).
\label{eq9}
\end{eqnarray}

To compute $D_{\rm ODP}$, we choose an initial electronic state $\bf{k}$, an optical phonon branch and phonon states $\bf{q}$ on that branch along a high-symmetry line (i.e. $\Gamma$-X), and for those $\bf{q}$ phonon states we identify the final electronic state having $\bf{k'=k+q}$, for which we compute $M_{mn}^\nu (\bf{k},\bf{q})$. Again, commonly for simplicity, transport studies, especially device simulations, consider a single value for the deformation potentials for all transitions from a given initial electronic band to a given final band (in general it can be the same or a different band), and a single dominant phonon energy \cite{Samsonidze2018}. This is justified as we will see below, since both $M_{mn}^\nu (\bf{k},\bf{q})$ and the phonon branch energy are relatively constant. The corresponding scattering rate $|S_{\bf{k,k'}}^{\rm{ODP}}|$ can then be computed as \cite{Lundstrom2000,Neophytou2020}:
\begin{eqnarray}
|S_{\bf{k,k'}}^{\rm{ODP}}|=\frac{\pi D_{\rm{ODP}}^2} {2 \rho \omega} (N_{\omega} + \frac{1}{2} \mp \frac{1}{2})  \textsl{g}_{\bf{k'}},
\label{eq10}
\end{eqnarray}
where $\omega$ is the frequency of the optical phonons near the $\Gamma$-point, which is considered to be constant. $N_\omega$ is the phonon Bose-Einstein statistical distribution and the “$+$” and “$-$” sign indicate the emission and absorption processes, respectively.

\section{Deformation potentials}%

Below, using the method described, we use Si as an example to show how to derive the deformation potentials and compute the transport properties. The electronic band structure, phonon dispersion, and e-ph coupling matrix elements are calculated from DFT and DFPT using the QUANTUM ESPRESSO package \cite{Giannozzi2009}. The optimized norm-conserving Vanderbilt (ONCV) \cite{Hamann2013} pseudopotential is used for Si under the generalized gradient approximation (GGA) with the Perdew-Burke-Ernzerhof (PBE) \cite{Perdew1996} functional. Since the outcomes of DFT calculations could depend on the choice of pseudopotentials and exchange-correlation functionals, we have also performed comparisons using three more sets of functionals, and we briefly discuss the outcomes later on in the discussion section and Appendix B. The 12$\times$12$\times$12 and 18$\times$18$\times$18 Monkhorst-Pack $\bf k$ meshes are used for structure relaxation and electronic band structure calculations, respectively. The cutoff energy of plane waves is set to 35 Ry. All of the parameters have been tested to be sufficient in obtaining converged results. The relaxed lattice constants is 5.479 \AA, indicating a slight 0.88\% overestimation with respect to the available experimental value of 5.431 \AA\ \cite{Robert1996}, which is the general tendency of GGA \cite{Stampfl2001}. The EPW package \cite{Ponce2016} is used to perform Wannier function interpolation for the e-ph coupling matrix elements. Initial coarse 12$\times$12$\times$12 $\bf k$ and 6$\times$6$\times$6 $\bf q$ meshes are used. The denser $\bf k$-grid for Si is needed in order to obtain a good Wannier interpolation of the conduction bands \cite{Ponce2018}, since the minimum does not reside on a high-symmetry point. 

Figure \ref{fig2} shows the electronic band structure and phononic spectrum for Si. Using these dispersions, we compute the deformation potentials for acoustic and optical phonons for both holes and electrons. To compute the e-ph matrix elements, we set the initial electronic state at the VBM for holes and at the CBM for electrons. We then consider acoustic and optical phonons of wave vector $\bf{q}$ (long wavelengths) along a high-symmetry line. The final state in the scattering process then has a wave vector $\bf{k+q}$, and is also taken along a high-symmetry line in the Brillouin zone. In the case of optical phonons we calculate the zero-order deformation potential, whereas in the case of acoustic phonons we calculate the first-order deformation potential. The red dots and line regions in Fig. \ref{fig2} indicate the electronic and phononic triplets that take part in the calculation, with the larger central dots in Fig. \ref{fig2}(a) being the initial states in the CBM and VBM, respectively, whereas the red lines being the final electronic states involving the phonon states indicated in Fig. \ref{fig2}(b).

\subsection{Hole-phonon coupling matrix elements}

\begin{figure}[tbp]
\includegraphics[width=0.48\textwidth]{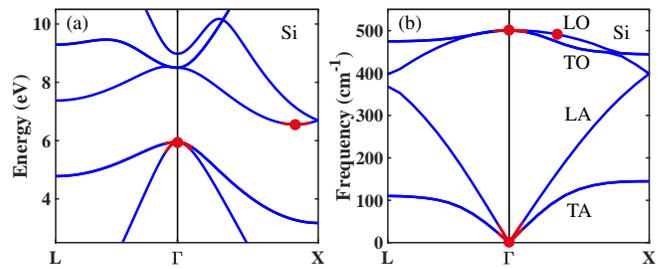}
\caption{(a) The electronic structure and (b) phononic spectrum of Si. In (a) we depict with large red dots for the initial electronic state at the VBM for holes and at the CBM for electrons, and red line segments for the final electronic states. In (b) the large red dots and red line segments indicate the corresponding phonons that are involved in the transitions for intravalley and $g$-type intervalley scatterings. The phonons for $f$-type scattering are not shown here as they are not located at the L-$\Gamma$-X path.}
\label{fig2}
\end{figure}

\begin{figure}[tbp]
\includegraphics[width=0.48\textwidth]{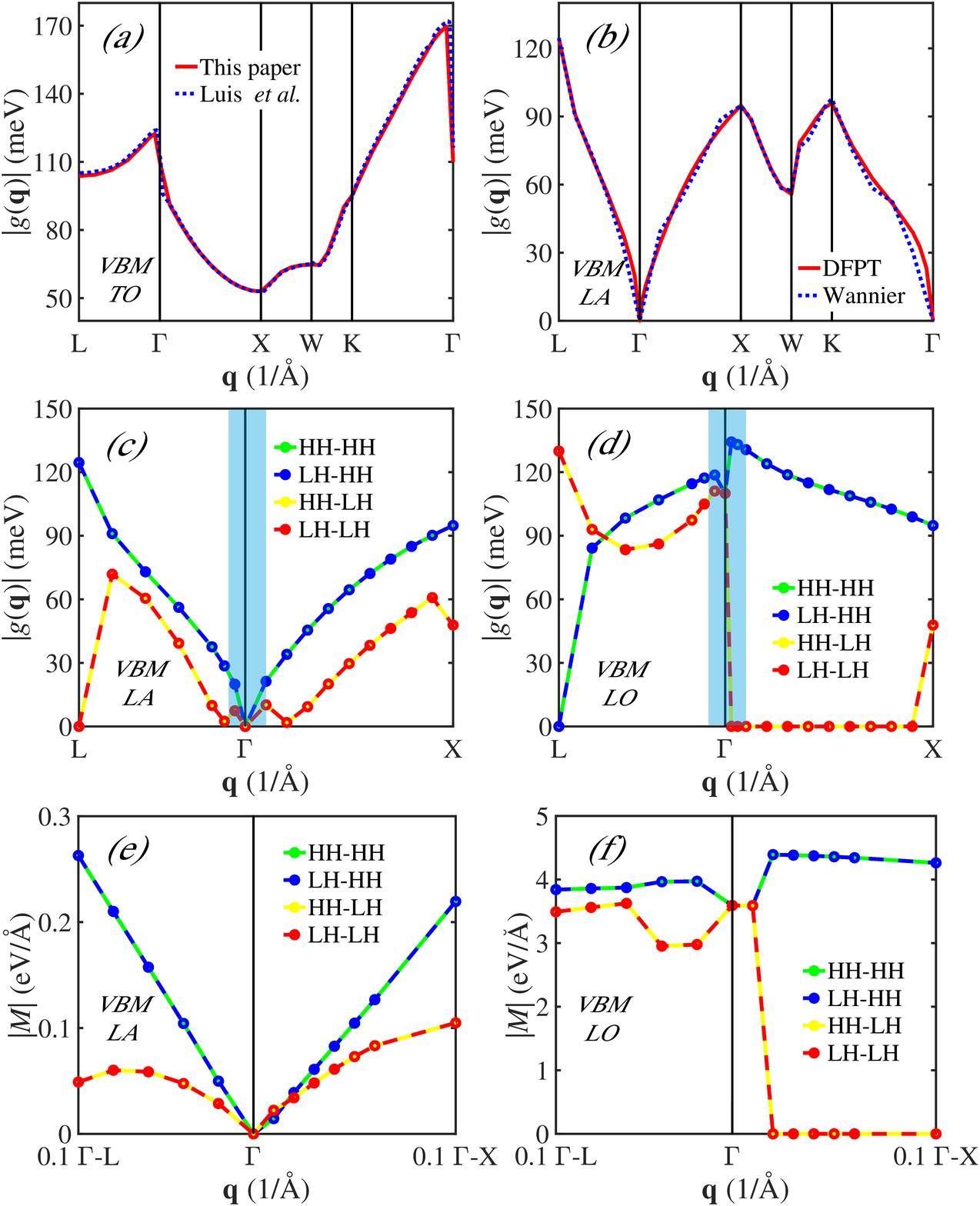}
\caption{Matrix elements [(a)-(d)] and $M_{mn}^\nu (\bf{k},\bf{q})$ [(e)-(f)] for the valence band of Si vs the phonon wave vector $\bf{q}$. The initial electronic state $\bf{k}$ is at the VBM for holes ($\Gamma$ point), whereas the corresponding final states are $\bf{k'= k + q}$. (a) Our calculated matrix elements for the TO phonon mode (solid line) compared to Luis’s work (dotted line) \cite{Agapito2018}. (b) Matrix elements for the LA phonon mode (solid line), compared to that using Wannier interpolations (dotted line). (c) Calculated matrix elements for the LA phonon with a separate combination of transitions from the heavy hole (HH) or light hole (LH) bands as initial/final states. The blue regions near the $\Gamma$ point indicate the most relevant matrix elements for the scattering processes. The corresponding four combinations of transitions between HH and LH are indicated. (d) Same as (c) but for the LO phonon mode. (e) $M_{mn}^\nu (\bf{k},\bf{q})$ of the VBM for the LA and (f) LO phonon modes. The $\bf{q}$ vector range is the zoom around the blue regions of (c) and (d).}
\label{fig3}
\end{figure}

We first take the e-ph matrix elements with regards to the coupling of holes to the transverse optical mode (TO) $g_{\rm{VBM,TO}}$ as an example, where the same calculation can be found in the literature and a direct comparison can be performed \cite{Agapito2018}. The initial electronic state is located at the VBM and the TO mode is considered (labeled in Fig. \ref{fig2}). Due to energy/momentum conservation \cite{Lundstrom2000}, only phonons of small $\bf q$, around the $\Gamma$ point of the phonon spectrum [Fig. \ref{fig2}(b)], take part in scattering processes involving states around the valence band maximum (a few $k_BT$). This is the case for both hole-acoustic and hole-optical phonon scattering (in the case of optical phonons, emission/absorption processes result in energy changes by $\sim$61 meV as shown later on). Our calculated e-ph matrix elements agree well with those found in the literature \cite{Agapito2018}, as shown in Fig. \ref{fig3}(a), which shows the $g_{\rm{VBM,TO}}$ as a function of the phonon ${\bf{q}}$ vector along high-symmetry lines for our calculation (red solid line) and literature data (blue dashed line). Note that here we compute the matrix elements involving phonons from the entire phonon spectrum (although phonons away from the $\Gamma$ point do not participate in scattering events). 

Since the VBM of Si is located at the $\Gamma$ point which is the center of the Brillouin zone of a cubic semiconductor, the deformation potential constant for acoustic phonons is a second-rank tensor with cubic symmetry. It has a diagonal form with equal diagonal elements and therefore can be treated as a scalar quantity. Its trace is non-zero for longitudinal phonons (referred to as square term $D_{\rm{ADP}}^2q^2$ as well \cite{Jacoboni2010}), but it vanishes for transverse modes \cite{Jacoboni2010,Peter2010} for symmetry reasons at the $\Gamma$ point. The detailed proof for acoustic phonons can be found in Ref. \cite{Peter2010}. Essentially, the LA mode contributes to volume changing deformations (dilatation component) at first order and accountable scattering rates, whereas its shear components are usually less important \cite{Peter2010}. Therefore we can neglect the effect of the shear strain and treat LA phonons as a change in the volume of the crystal, which gives rise to a perturbing potential that shifts the electronic band energy. On the other hand, the TA modes contain only shear waves and contribute to shear and nonvolume changing deformations only (and at first order their effect on volume change can be ignored), which introduce scattering rates at second order. Thus, the dominant acoustic mode is only the LA for the VBM, whereas the TA plays a secondary role. For optical phonons, the scattering rates can be derived similarly to acoustic phonons by replacing the squared factor $D_{\rm{ADP}}^2q^2$ with a squared optical coupling constant $D_{\rm{ODP}}^2$ \cite{Jacoboni2010}. Thus, for hole scattering in Si (with the VBM at the $\Gamma$ point) we consider only the longitudinal phonons and ignore the transverse phonons, though the coupling matrices of TA and TO modes are not zero (see Appendix C). First-principles calculations, e.g., using the EPW software for the strength of scattering rates from different phonon branches in the case of GaAs, where the VBM and CBM are both at the $\Gamma$ point, indeed show that the scattering rates are dominated by the LO and LA phonons \cite{Ma2018}, rather than the transverse phonons, which are associated with significantly weaker (at second order) scattering rates. On the other hand, for electrons in Si (where the CBM is not at the $\Gamma$ point) we need to consider all LA, TA, LO, and TO phonons, all of which contribute to the scattering rates \cite{Qiu2015}.

To derive the acoustic and optical deformation potentials for holes, we need to calculate the e-ph matrix elements $g_{\rm VBM,LA}$ and $g_{\rm VBM,LO}$ with the initial electronic state $\bf{k}$ located at the VBM. The phonon mode states have momentum $\bf{q}$ which results in the final electronic states having momentum $\bf{k+q}$ since momentum conservation is enforced. Figure \ref{fig3}(b) shows the $g_{\rm VBM,LA}$ matrix elements over the high-symmetry $\bf{q}$ directions from DFPT calculations. It is compared to the method where Wannier interpolation is performed, indicating excellent agreement. Thus, we use Wannier interpolation to accelerate the calculation of matrix elements. Note that the Wannier interpolation method might not be crucial in the case of Si holes, since the VBM is located at a high-symmetry point. If the initial electronic states are located at nonhigh-symmetry points, e.g., the case for the CBM of Si, a very dense $\bf{k}$ mesh is required to include the initial electronic state in the DFPT calculation of matrix elements directly. To avoid a large number of calculations, one can use the maximally localized Wannier functions to interpolate the e-ph matrix elements. Wannier interpolation allows the free choice of the initial electronic state and can more conveniently consider various $\bf{q}$ paths, while the actual DFT and DFPT calculations are still performed on coarse $\bf{k}$ and $\bf{q}$ meshes.

\begin{figure}[tbp]
\includegraphics[width=0.4\textwidth]{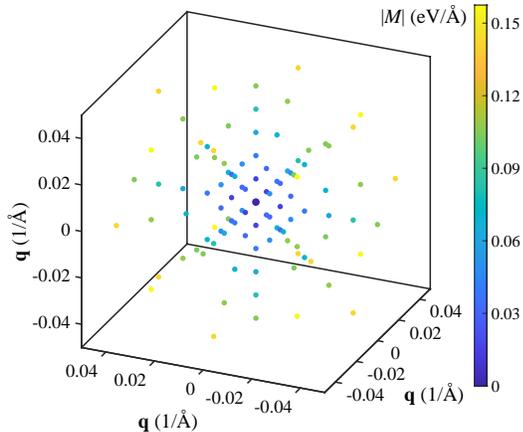}
\caption{3D view of $M_{mn}^\nu (\bf{k},\bf{q})$ for LA phonons for scattering into the HH along different $\bf{q}$ vectors.}
\label{fig4}
\end{figure}

Previous works show that spin-orbit coupling (SOC) has a significant effect on the mobility of holes, especially at low temperatures \cite{Ma2018}, which is related to the contributions of band splitting on low-energy holes. At and above room temperature, the mobilities of holes with and without SOC are similar \cite{Ponce2018}. Therefore, to consider the effect of band splitting, we include SOC for the band structures that we use within the BTE calculations later on. However, we find that SOC has little effect on the deformation potentials (see Appendices D and E). Therefore, in the derivation of deformation potentials, the DFT and DFPT calculations are carried out without SOC, and thus the VBM consists of three degenerate bands at the $\Gamma$ point. Away from the minimum, there are two doubly degenerated valance bands (heavy hole, HH) and one singly degenerated band (light hole, LH). We consider the matrix elements that arise for the different combinations between the initial and final electronic states residing at the HH or the LH, i.e., HH-LH indicates the matrix elements from an initial HH band to a final LH band. For a given initial state ${\bf{k}}$ on band $n$, there are two final states ${\bf{k'}}$ on each of the HH bands and one final state on the LH band. Note that we also find that SOC does not affect the wave functions in the nondegenerate case for electrons.

\begin{table*}
\caption{Intravalley acoustic deformation potential (eV) and intravalley optical deformation potential (eV/\AA) for LA and LO phonons along different high-symmetry directions of ${\bf q}$, for both the $\Gamma \rightarrow \Gamma + {\bf q}$ and $\Gamma - {\bf q}/2 \rightarrow \Gamma + {\bf q}$/2 transitions.}
\begin{ruledtabular}
\begin{tabular}{lcccccc}
 & & & X & L & K & Average \\
\hline
$\Gamma \rightarrow \Gamma + {\bf q}$ & LH-LH & LA & 0.9145 & 0.4931 & 0.1470 & 0.5270 \\
& & LO & 0 & 3.4921 & 0 & 1.937 \\
& LH-HH & LA & 1.9141 & 2.6473 & 2.3147 & 2.3398 \\
& & LO & 4.2620 & 3.8414 & 4.2396 & 4.1265 \\
& HH-HH & LA & 1.9141 & 2.6473 & 2.3147 & 2.3398 \\
& & LO & 4.2620 & 3.8414 & 4.2396 & 4.1265 \\
& HH-LH & LA & 0.9145 & 0.4931 & 0.1470 & 0.5270 \\
& & LO & 0 & 3.4921 & 0 & 1.937 \\
\hline
$\Gamma - {\bf q}/2 \rightarrow \Gamma + {\bf q}$/2 & LH-LH & LA & 2.1754 & 1.9232 & 1.6028 & 1.8482 \\
& & LO & 0 & 6.3378 & 0 & 3.5156 \\
& LH-HH & LA & 0 & 0 & 0 & 0 \\
& & LO & 0 & 0 & 0 & 0 \\
& HH-HH & LA & 2.1300 & 3.1822 & 2.3304 & 2.5825 \\
& & LO & 5.3672 & 4.7461 & 5.2490 & 5.1279 \\
& HH-LH & LA & 0 & 0 & 0 & 0 \\
& & LO & 0 & 0 & 0 & 0 \\
\end{tabular}
\end{ruledtabular}
\label{table1}
\end{table*}

The calculated e-ph matrix elements $g_{\rm VBM,LA}$ along the high-symmetry paths $\Gamma$-X, $\Gamma$-L, and $\Gamma$-K for the phonon wave vectors $\bf{q}$ are shown in Fig. \ref{fig3}(c). It is interesting to see that the matrix elements describing transitions with HH as the final state, i.e. HH-HH and LH-HH, are identical. The same occurs for transitions that have final states on the LH band, i.e. HH-LH and LH-LH. The reason is that we have picked the $\Gamma$ point as the initial point for all the matrix elements. Due to the T$_{2 \rm g}$ band symmetry at that point, which belongs to the O$_{\rm h}$ group, the initial wavefunctions of all three states on the two HH and the LH bands are the same. Thus, differences in the matrix elements $\langle {\psi_{m{\bf{k+q}}}}({\bf{r}})|{\delta _{\nu {\bf{q}}}}V({\bf{r}})|{\psi _{n{\bf{q}}}}({\bf{r}}) \rangle$ for the same initial state are only determined by the final states, i.e. whether a carrier scatters into the HH or the LH. This allows for only two independent values for the matrix elements, scattering into the HH or the LH.

The acoustic phonon matrix elements for scattering into the HH are larger compared to those representing scattering into the LH. The $M_{mn}^\nu (\bf{k},\bf{q})$ to HH [Fig. \ref{fig3}(e)] increases linearly with $\bf{q}$, indicating stronger transition rates and scattering for the scattering events into the HH, whereas the acoustic phonon scattering into the LH is a weaker process with smaller $M_{mn}^\nu (\bf{k},\bf{q})$. The acoustic $M_{mn}^\nu (\bf{k},\bf{q})$ for scattering into the LH do not increase linearly. Only for a small $\bf{q}$ vector, e.g., less than one-tenth of half Brillouin zone, one can consider that a slope can be taken and the deformation potential extracted. The shape of the matrix elements is not even isotropic, bending more in the $\Gamma$-L direction. 

With regards to the optical phonon scattering with the LO phonons, the matrix elements for scattering from HH into HH are nearly constant and slightly anisotropic along the high-symmetry $\Gamma$-X, $\Gamma$-L, and $\Gamma$-K paths (the latter not shown). The optical $M_{mn}^\nu (\bf{k},\bf{q})$ with final states on the LH band is non-zero along $\Gamma$-L, but falls to zero along the high-symmetry $\Gamma$-X and $\Gamma$-K paths, which is related to the symmetry along different directions \cite{Lax1961}. This again signals even weaker scattering processes for scattering into the LH. 

As the matrix elements generally depend on the eigenvectors of the phonon state, which are in general anisotropic, for simplicity we define the averaged deformation potential along all the directions, i.e., $\Gamma$-X, $\Gamma$-L, and $\Gamma$-K high-symmetry directions shown in Fig. \ref{fig4}, as
\begin{eqnarray}
D = \sqrt{\frac{n_{\Gamma \text{-} \rm{X}} D_{\Gamma \text{-} \rm{X}}^2+n_{\Gamma \text{-} \rm{L}} D_{\Gamma \text{-} \rm{L}}^2+ n_{\Gamma \text{-} \rm{K}} D_{\Gamma \text{-} \rm{K}}^2} {n_{\Gamma \text{-} \rm{X}}+n_{\Gamma \text{-} \rm{L}}+n_{\Gamma \text{-} \rm{K}}}}.
\label{eq11}
\end{eqnarray}
For a face-centered cubic (FCC) lattice, the number of the equivalent crystallographic orientations $n_{\Gamma \text{-} \rm{X}}$, $n_{\Gamma \text{-} \rm{L}}$, and $n_{\Gamma \text{-} \rm{K}}$ are 6, 8, and 12, respectively. In the case of the acoustic phonons, we take the averaged slope of $M_{mn}^\nu ({\bf{k},\bf{q}})$, whereas in the case of optical phonons, they are given by the averaged $M_{mn}^\nu ({\bf{k},\bf{q}})$.  

In addition to the initial state residing on the $\Gamma$, we also consider the $M_{mn}^\nu (\bf{k},\bf{q})$ along high-symmetry directions for $\Gamma - {\bf q}/2 \rightarrow \Gamma + {\bf q}$/2 transitions, where both the initial and final states are around the ${\Gamma}$. We consider the $n_{\Gamma \text{-} \rm{X}}$, $n_{\Gamma \text{-}  \rm{L}}$, and $n_{\Gamma \text{-}  \rm{K}}$ directions. The deformation potentials along different directions and the averaged value calculated using Eq. (\ref{eq11}) are shown in Table I. We can find that the LA and LO deformation potentials are different for the transitions $\Gamma - {\bf q}/2 \rightarrow \Gamma + {\bf q}$/2 compared to $\Gamma \rightarrow \Gamma + {\bf q}$. We can combine the deformation potentials for degenerated bands as the total one: \cite{Sjakste2015}
\begin{eqnarray}
D_{\rm tot} = \sqrt{\sum_{nm} (D_{nm})^2}.
\label{eq12}
\end{eqnarray}
The so-calculated $D_{\rm tot}$ for LA and LO are 5.80 eV and 10.65 eV/\AA, respectively, for $\Gamma \rightarrow \Gamma + {\bf q}$, and 5.49 eV and 10.84 eV/\AA, respectively, for $\Gamma - {\bf q}/2 \rightarrow \Gamma + {\bf q}$/2 transition. The difference of deformation potentials with the same $\bf q$ but different initial $\bf k$ points, is due to the symmetry and wavefunctions of the degenerate bands. Although the strength of the individual processes can differ if the initial state is around, rather than on $\Gamma$, still, the overall valence band deformation-potential values are comparable. 

Compared to the total deformation potential method above, another more physically clear method to define the overall deformation potential can be employed by using the largest deformation potential upon rotation of the wavefunctions in the subspace of the degeneracies. The ideal is that the global wave function of degenerate states is a linear combination of all individual states, and the matrix elements need to include the global wave function, and not individual processes for all separate/individual states/eigenvectors. Numerically, the overall deformation potential can be obtained by constructing a tensor out of the deformation potentials of the different degenerate bands and taking the largest singular value (or all if the others contribute significantly) \cite{Vandenberghe2015} (more details can be found in Appendix D). The required wave functions rotation reflects in the unitary matrices within the singular value decomposition. We have tested the so-calculated deformation potentials for two different cases around the $\Gamma$ (see Appendix D). For the transitions of $\Gamma \rightarrow \Gamma + {\bf q}$, the largest singular values are 5.80 eV and 10.65 eV/\AA\ for LA and LO, respectively. For the transitions of $\Gamma - {\bf q}/2 \rightarrow \Gamma + {\bf q}$/2, they are 5.17 eV and 10.26 eV/\AA\ for LA and LO, respectively. Each of them accounts for the nine matrix elements $\langle {\psi_{m{\bf{k+q}}}}({\bf{r}})|{\delta _{\nu {\bf{q}}}}V({\bf{r}})|{\psi _{n{\bf{q}}}}({\bf{r}}) \rangle$ for all possible initial and final states. Using the averaged values of the two transitions, the acoustic and optical deformation potentials for holes in Si result to $D_{\rm{ADP}}$ = 5.48 eV and $D_{\rm{ODP}}$ = 10.45 eV/\AA.

\subsection{Electron-phonon coupling matrix elements}

The CBM in Si is formed from six equivalent valleys, located along the $\Gamma$-X direction in the Brillouin zone, at $\sim$84\% towards the X point. To compute the coupling matrix elements in the conduction band, we use one of the six equivalent ellipsoids as the initial state, $\bf k$ = (0, 0, 0.8375), and final states within the same ellipsoid (intravalley transitions) and in the other five ellipsoids (intervalley transitions). 

\subsubsection{Intravalley transitions}

\begin{figure*}[tbp]
\includegraphics[width=0.8\textwidth]{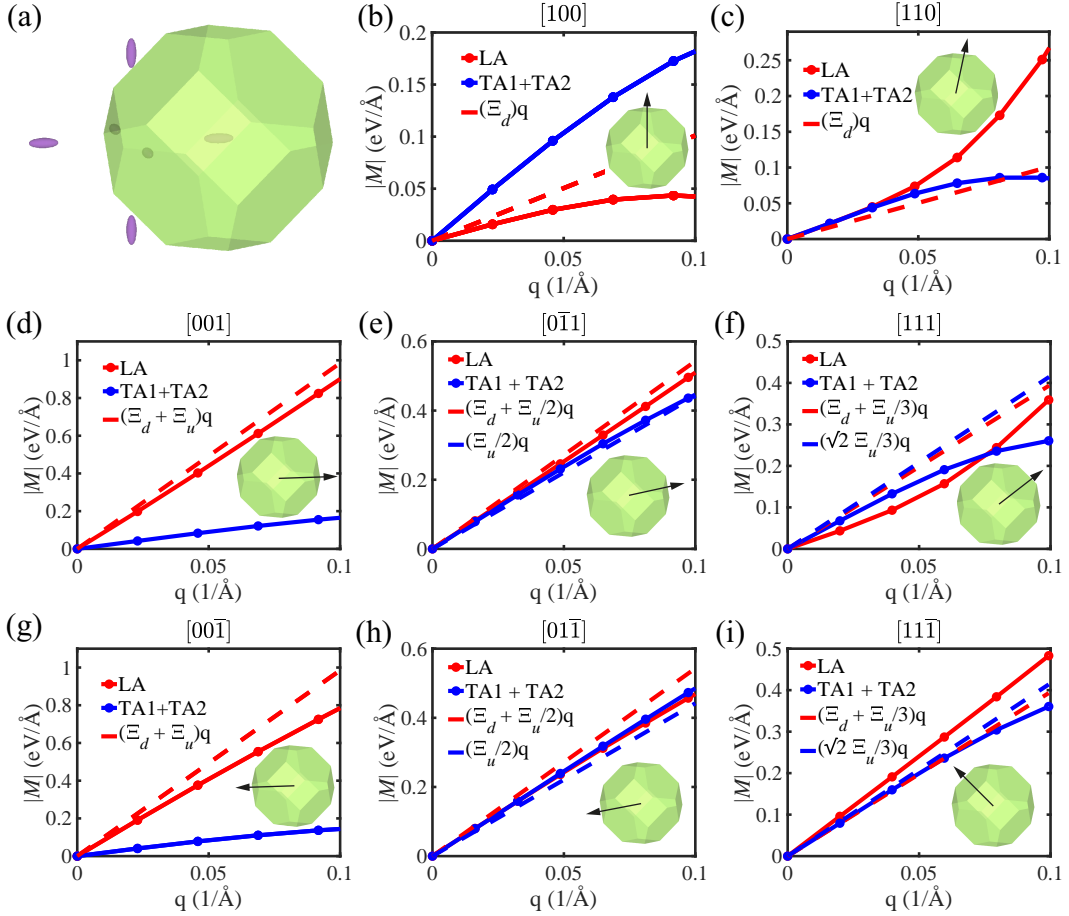}
\caption{Calculated intravalley electron-phonon coupling matrix elements for Si as a function of the phonon momentum $|{\bf{q}}|$, for transitions via LA (red solid lines) or TA (blue solid lines) phonons. For $\bf q$ $\rightarrow$ 0, the relation between the $M_{mn}^\nu (\bf{k},\bf{q})$ and the dilatation and uniaxial deformation potentials for LA phonons and the sum of the two TA phonons (see Table II) are shown with the dashed lines. (a) The first Brillouin zone of $\bf q$ vectors, where the center is located at the CBM of the electronic Brillouin zone, at $\bf k$ = (0, 0, 0.8375). [(b)-(i)] Electron-phonon $M_{mn}^\nu (\bf{k},\bf{q})$ (or the zero-order deformation potential) in the direction of $\bf q$ as shown in the insets.}
\label{fig5}
\end{figure*}

For the intravalley electron-phonon scattering, we focus first on the acoustic deformation potentials. The direct way to calculate the averaged deformation potentials for LA and TA is using the root mean square of the slopes of the $M_{mn}^\nu (\bf{k},\bf{q})$ for LA and TA, as shown by the solid red and blue lines in Fig. \ref{fig5}. 
If we define $\theta$ as the angle between the phonon wave vector and the longitudinal axis of the conduction band valley, the slopes for LA with $\theta = 0, 0.25\pi, 0.304\pi, 0.5\pi, 0.75\pi, 0.696\pi, \pi$ [the different directions shown in Figs. \ref{fig5}(b)-\ref{fig5}(i)], are 8.98, 5.07, 2.63, 1.01, 4,91, 4.83, 8.33 eV, respectively. For TA, the slopes are 1.83, 4.87, 3.40, 1.75, 4.90, 3.99, 1.77 eV, respectively. The so-calculated averaged deformation potentials for LA and TA are 5.75 eV and 3.48 eV, respectively. This method requires a large number of directions so that an accurate averaged value is obtained.

A more general way to compute the averaged deformation potentials is considering the symmetry properties of Si, which allows a reduction to just two independent potentials. These are the dilatation deformation potential $\Xi_d$ and the uniaxial shear deformation potential $\Xi_u$. In contrast to holes with isotropic deformation potentials, the deformation potential of electrons has a general angular dependence as \cite{Herring1956}:
\begin{eqnarray}
\Xi_{\rm LA}(\theta) = \Xi_d + \Xi_u {\rm cos}^2 \theta,
\label{eq13}
\end{eqnarray}
\begin{eqnarray}
\Xi_{\rm TA}(\theta) = \Xi_u {\rm sin} \theta{\rm cos} \theta.
\label{eq14}
\end{eqnarray}
It should be mentioned that both transverse modes are incorporated here. To use those, we consider eight directions, $[100]$, $[110]$, $[001]$, $[00\overline 1]$, $[0\overline 1 1]$, $[01\overline 1]$, $[111]$, and $[11\overline 1]$. We then compute the LA and TA electron-phonon coupling matrices along these high-symmetry $\bf q$ paths [Figs. \ref{fig5}(b)-\ref{fig5}(i)], which can be expressed in terms of $\Xi_d$ and $\Xi_u$ \cite{Herring1956}, as derived and listed in Table II. From these relations and the LA and TA matrix elements in Figs. \ref{fig5}(b)-\ref{fig5}(i), along different directions, several values for $\Xi_d$ and $\Xi_u$ can be extracted. These $\Xi_d$/$\Xi_u$ values extracted for each direction are in general similar, but some differences for different directions can occur. Thus, we compute the averaged deformation potential, with all $\bf q$ directions (with a nonzero linear term) are weighted equally, i.e., $\Xi_d$ is calculated by deriving the slope of $|M|$ for the LA with $\bf q$ $\rightarrow$ 0 along $[100]$ and $[110]$, while $\Xi_u$ is calculated by deriving the slope of $|M|$ for the TA with $\bf q$ $\rightarrow$ 0 along $[0\overline 1 1]$, $[01\overline 1]$, $[111]$, and $[11\overline 1]$. This procedure yields the values of $\Xi_d$ = 1.01 eV and $\Xi_u$ = 8.84 eV.

Figure \ref{fig5}(a) shows the Brillouin zone of phonon $\bf{q}$ for which the center overlaps with one of the CBM ellipsoids, (0, 0, 0.8375). The electron-phonon $M_{mn}^\nu (\bf{k},\bf{q})$ for different phonon polarizations, for both the LA/TA and $\Xi_d$/$\Xi_u$ extracted elements, are shown in Figs. \ref{fig5}(b)-\ref{fig5}(i) for different $\bf{q}$-vector directions as illustrated in the sub-figures and noted. In most cases, the lines from the two methods agree very well. Some discrepancies can be observed along the $[100]$ and $[110]$ directions, and somewhat smaller in the $[111]$ and $[11\overline 1]$ directions. In those cases, the slopes of the coupling matrices at $|{\bf q}| \rightarrow 0$ are different from those constructed using the $\Xi_d$/$\Xi_u$ expressions in Table II. The calculated coupling matrices are not linear at large $|{\bf q}|$ values, which increases the mismatch.

\begin{table}
\caption{The expressions for $|M|$/$|\bf q|$ for the longitudinal and transverse acoustic phonons along different high-symmetry $\bf q$ directions for the CBM at $\bf k$ = (0, 0, 0.8375) of a cubic material. $\Xi_d$ and $\Xi_u$ represent dilatation and uniaxial deformation potentials, respectively.}
\begin{ruledtabular}
\begin{tabular}{ccc}
$\bf q$ & LA & TA1 + TA2\\
\hline
$[100]$ & $\Xi_d$ & 0 \\
$[110]$ & $\Xi_d$ & 0 \\
$[001]$ & $\Xi_d$ + $\Xi_u$ & 0 \\
$[00\overline 1]$ & $\Xi_d$ + $\Xi_u$ &  0\\
$[0\overline 1 1]$ & $\Xi_d$ + $\Xi_u$/2 & $\Xi_u$/2 \\
$[01\overline 1]$ & $\Xi_d$ + $\Xi_u$/2 & $\Xi_u$/2 \\
$[111]$ & $\Xi_d$ + $\Xi_u$/3 & $\sqrt{2}$ $\Xi_u$/3 \\
$[11\overline 1]$ & $\Xi_d$ + $\Xi_u$/3 & $\sqrt{2}$ $\Xi_u$/3 \\
\end{tabular}
\end{ruledtabular}
\label{table2}
\end{table}

Previous calculations have shown that the influence of the angular {\bf q}-dependence on electron transport is relatively small \cite{Mizuno1993}. Thus, we consider and extract the isotropically averaged intravalley deformation potentials for LA and TA phonons as an approximation, derived from the average of the integrals of squared $\Xi_{\rm LA}(\theta)$ and $\Xi_{\rm TA}(\theta)$ as:
\begin{eqnarray}
D_{\rm LA}^2 = \Xi_d^2 + \Xi_d \Xi_u + \frac{3}{8} \Xi_u^2,
\label{eq15}
\end{eqnarray}
\begin{eqnarray}
D_{\rm TA}^2 = \frac{1}{8} \Xi_u^2.
\label{eq16}
\end{eqnarray}
Using Eqs. (\ref{eq15}) and (\ref{eq16}), the deformation potentials for LA and TA phonons turn out to be 6.27 eV and 3.13 eV, respectively, which are similar to the values extracted directly from averaging the LA and TA matrix elements above. 

\begin{figure*}[tbp]
\includegraphics[width=0.8\textwidth]{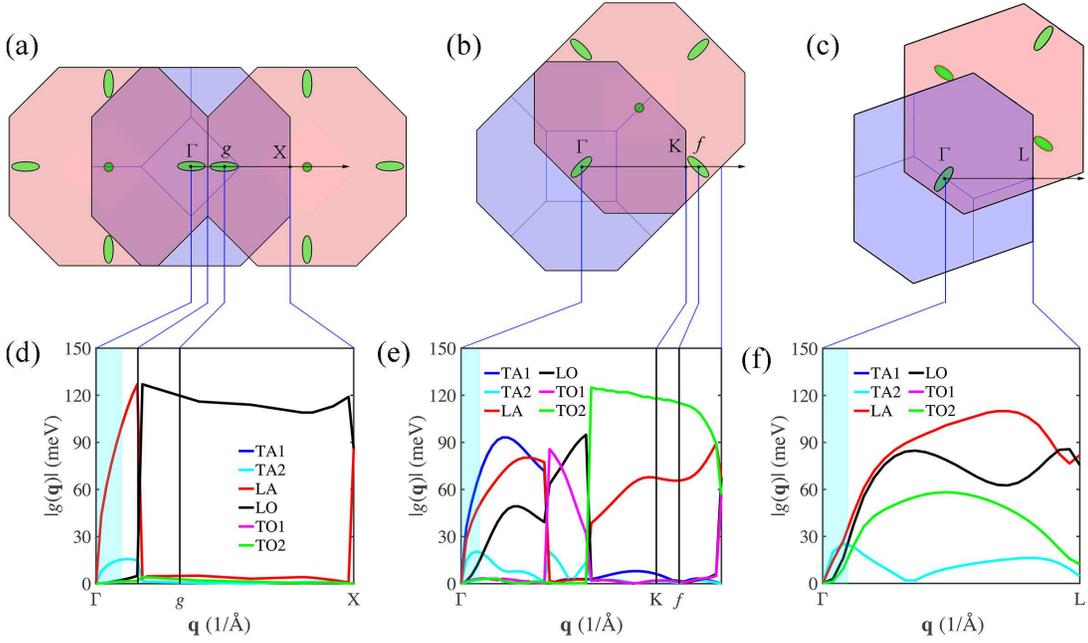}
\caption{(a)-(c) 2D cross-section view of the first Brillouin zone for electrons (red) and phonons (blue), with the 6 CBM depicted by the green ellipsoids. The corresponding matrix elements for electrons vs the phonon wavevector $\bf{q}$ along (d) $\Gamma$-X direction, also indicating the $g$-type intervalley process, (e) $\Gamma$-K direction, also indicating the $f$-type process, and (f) $\Gamma$-L direction, with initial electronic state $\bf{k}$ at the CBM, where the blue zone within 1/10 of half Brillouin zone is considered for the derivation of the coupling matrix $M$ for acoustic phonons.}
\label{fig6}
\end{figure*}

The scattering rates for electrons can be calculated as
\begin{eqnarray}
|S_{\bf{k,k'}}^{\rm{LA}}|=\frac{\pi}{\hbar} D_{\rm{LA}}^2 \frac{k_B T}{\rho v_l^2}\textsl{g}_{\bf{k'}},
\label{eq17} 
\end{eqnarray}
\begin{eqnarray}
|S_{\bf{k,k'}}^{\rm{TA}}|=\frac{\pi}{\hbar} D_{\rm{TA}}^2 \frac{k_B T}{\rho v_t^2} \textsl{g}_{\bf{k'}},
\label{eq18}
\end{eqnarray}
where $v_l$ = 9.04 km/s \cite{Jacoboni1977} and $v_t$ = 5.34 km/s \cite{Jacoboni1977} are velocities of the longitudinal and transverse acoustic phonons, respectively. The expressions for the scattering rates due to the interaction with LO and TO phonons can be derived similarly to those for acoustic phonons, and read as \cite{Cao2018}
\begin{eqnarray}
|S_{\bf{k,k'}}^{\rm{LO}}|=\frac{\pi D_{\rm{LO}}^2} {2 \rho \omega_{\rm{LO}}} (N_{\omega} + \frac{1}{2} \mp \frac{1}{2})  \textsl{g}_{\bf{k'}},
\label{eq19}
\end{eqnarray}
\begin{eqnarray}
|S_{\bf{k,k'}}^{\rm{TO}}|=\frac{\pi D_{\rm{TO}}^2} {2 \rho \omega_{\rm{TO}}} (N_{\omega} + \frac{1}{2} \mp \frac{1}{2})  \textsl{g}_{\bf{k'}}.
\label{eq20}
\end{eqnarray}
In this case for the intravalley electron-phonon scattering of Si, the optical matrix elements go to zero as the phonon wave vector approaches zero, dictated by symmetry selection rules \cite{Lax1961}. Indeed, the zero-order deformation potentials for LO and TO modes are small around the $\Gamma$-point, as shown in the blue highlighted regions in Fig. \ref{fig6}. Small $\bf q$ vectors correspond to intravalley transitions, and it is well known that only acoustic phonons contribute to that, while optical phonons in Si conduction bands only cause intervalley transitions, as we elaborate below. Although we observe some small values, especially along the $\Gamma$-L direction, intravalley transitions caused by optical phonons are typically neglected, and we take $D_{\rm ODP}$ = 0 eV/\AA\ for electrons. 

In addition, it is customary in the literature to use only the longitudinal mode for scattering, for example in Monte Carlo models \cite{Canali1975,Lundstrom2000}, in which case the contributions from LA and TA modes are combined as $D_{\rm ADP}^2 = D_{\rm{LA}}^2 + (v_l^2/v_t^2)D_{\rm{TA}}^2$. The overall value is then extracted to be $D_{\rm ADP}$ = 8.21 eV, which also agrees well with what is routinely employed. 

\subsubsection{Intervalley transitions}

Two types of intervalley scattering are possible in Si: One is referred to as the $g$-type process, which scatters a carrier from a given valley into one on the opposite side of the same axis, or its equivalent in the next Brillouin zone nearby; the other is the $f$-type process which scatters a carrier into one of the neighboring valleys on the perpendicular axes. Intervalley scattering requires very large changes in momentum, and requires phonons with wave vectors near the zone boundary to scatter electrons within the first, or even the second equivalent Brillouin zone, which can be either acoustic or optical phonons. Considering that away from the $\Gamma$ point the phonon momentum is both large and flat [see Fig. \ref{fig2}(b)], the intervalley scattering resembles the optical phonon scattering. Thus, we use the intervalley deformation potential $D_{\rm{IVS}}$ to characterize the strength of the scattering from the initial to the final valley, in the form of an optical deformation potential as
\begin{eqnarray}
V_{\rm{e\text{-}ph}} = D_{\rm{IVS}} \bf{u},
\label{eq21}
\end{eqnarray}
where $D_{\rm{IVS}}$ (in units of energy per unit length, eV/m) can be calculated from:
\begin{eqnarray}
D_{\rm{IVS}} = M_{mn}^\nu ({\bf{k},\bf{q}}),
\label{eq22}
\end{eqnarray}
where the initial electronic state ${\bf{k}}$ is located at one of the CMB valleys, and the final electronic state ${\bf{k'=k+q}}$ is located at another CBM valley. The corresponding intervalley scattering rate $|S_{\bf{k,k'}}^{\rm{IVS}}|$ can then be computed using Eq. (\ref{eq10}) for the optical deformation potential scattering.

Figures \ref{fig6}(a)-\ref{fig6}(c) show the 2D cross-section views of the first Brillouin zone for electrons (red) and phonons (blue), where the six CBM are depicted by the green ellipsoids. The $g$-type and the $f$-type processes are illustrated in Figs. \ref{fig6}(a) and \ref{fig6}(b), respectively. The transitions involved in these two processes have different strengths. The corresponding matrix elements for electrons versus the phonon wavevector $\bf{q}$ with initial electronic state ${\bf k}$ at the CBM are shown in Figs. \ref{fig6}(d)-\ref{fig6}(f), respectively.

For the $g$ process we consider the initial and final states at (0, 0, 0.8375) and (0, 0, $-$0.8375) \cite{Long1960,Sinha2005}, i.e., the ellipsoids in the [001] and $[00\overline 1]$ directions which are located across each other. The matrix elements in the $\Gamma$-X direction in Fig. \ref{fig6}(d), indicate a strong intravalley scattering LA phonon contribution (red line) and a strong intervalley LO phonon contribution (black line) representing the $g$ process. The lines that connect to the Brillouin zone of Figure \ref{fig6}a clearly show that the $g$-process is associated with LO-caused transitions in the equivalent ellipsoid of the 2nd Brillouin zone in the extended $\Gamma$-X direction, i.e., 32.5\% of half Brillouin zone length away. It is also interesting to see that the matrix element associated with the LA mode collapses for transitions into the 2nd Brillouin zone, while that of the LO phonon has significant values only for transitions into the 2nd Brillouin zone. Note again that only the values around the $\Gamma$ point and the $g$-labelled point are of importance to transport, as only those involve energetically favorable electrons. 

\begin{table*}
\caption{The intravalley acoustic deformation potential $D_{\rm ADP}$ (eV) and intravalley optical deformation potential $D_{\rm ODP}$ (eV/\AA) with the corresponding phonon frequency (meV) for the holes, the dilatation deformation potential $\Xi_d$ (eV), uniaxial shear deformation potential $\Xi_u$ (eV), acoustic deformation potential (eV) for LA and TA phonons, and overall intravalley acoustic deformation potential $D_{\rm ADP}$ (eV), and the intervalley deformation potential (eV/\AA) for electrons in Si. Comparison is made with the deformation potentials and phonon frequencies found in the literature.}
\begin{ruledtabular}
\begin{tabular}{llll}
 & & This paper & Previous works\\
\hline
Holes, intravalley &  $D_{\rm{ADP}}$ (eV) & 5.48 &  5.0${^{\rm a,p}}$, 2.2\footnotemark[2], 3.1\footnotemark[14], 6.2${^{\rm q}}$, 7.12${^{\rm r}}$\\
 & $D_{\rm{ODP}}$  (eV/\AA) & 10.45 & 6\footnotemark[1], 5.0\footnotemark[2], 13.24${^{\rm e}}$, 9.05\footnotemark[14], 10.5\footnotemark[15]\\
 & Phonon frequency (meV) & 62.08 & 63\footnotemark[1] \\
Electrons, intravalley &  $\Xi_d$ (eV) & 1.01 & 1.1${^{\rm e}}$, 1.2${^{\rm k}}$, 1.13\footnotemark[12] \\
 & $\Xi_u$ (eV) & 8.84 & 10.5${^{\rm e}}$, 8.86${^{\rm f}}$, 8.47${^{\rm g}}$, 9.16${^{\rm h}}$, 9.29${^{\rm i}}$, 8.0${^{\rm j}}$, 8.6${^{\rm t}}$ \\
 & $D_{\rm{LA}}$ \rm (eV) & 6.27 & 6.39\footnotemark[13] \\
 & $D_{\rm{TA}}$ \rm (eV) & 3.13 & 3.01\footnotemark[13] \\
 & $D_{\rm{ADP}}$ (eV) & 8.21 & 9.5\footnotemark[1], 9.0\footnotemark[3], 7.8${^{\rm s}}$ \\
Electrons, intervalley & $g$-type, LO (eV/\AA) &  3.86 & 11\footnotemark[1], 3\footnotemark[3], 4.73\footnotemark[4] \\
 & Phonon frequency (meV) &  61.06 & 62.16\footnotemark[1], 62\footnotemark[3], 60\footnotemark[4] \\
 & $f$-type, LA (eV/\AA) &  1.83 & 2.0\footnotemark[1], 3.4\footnotemark[3], 2.51\footnotemark[4] \\
 & Phonon frequency (meV) & 46.67 & 47\footnotemark[1], 43\footnotemark[3], 47.73\footnotemark[4] \\
 & $f$-type, TO (eV/\AA) &  3.55 & 2.0\footnotemark[1], 4\footnotemark[3], 4.44\footnotemark[4] \\
 & Phonon frequency (meV) & 56.40 & 59\footnotemark[1], 54\footnotemark[3], 57.69\footnotemark[4] \\
\end{tabular}
\end{ruledtabular}
\footnotesize
${^{\rm a}}$Reference [\onlinecite{Lundstrom2000}];  ${^{\rm i}}$Reference [\onlinecite{Rieger1993}]; ${^{\rm b}}$Reference [\onlinecite{Ottaviani1975}];  ${^{\rm j}}$Reference [\onlinecite{Schmid1990}];${^{\rm c}}$Reference [\onlinecite{Canali1975}]; ${^{\rm k}}$Reference [\onlinecite{Yoder1994}]; ${^{\rm d}}$Reference [\onlinecite{Obukhov2009}]; ${^{\rm l}}$Reference [\onlinecite{Van1989}];\\
${^{\rm e}}$Reference [\onlinecite{Fischetti1996}];  ${^{\rm m}}$Reference [\onlinecite{Pop2004}]; ${^{\rm f}}$Reference [\onlinecite{Tserbak1993}]; ${^{\rm n}}$Reference [\onlinecite{Takeda1983}]; ${^{\rm g}}$Reference [\onlinecite{Friedel1989}]; ${^{\rm o}}$Reference [\onlinecite{Toshishige1995}]; ${^{\rm h}}$Reference [\onlinecite{Van1986}]; ${^{\rm p}}$Reference [\onlinecite{Jacoboni1983}];\\ 
${^{\rm q}}$Reference [\onlinecite{Dewey1993}]; ${^{\rm r}}$Reference [\onlinecite{Fischetti2003}]; ${^{\rm s}}$Reference [\onlinecite{Yu2008}]; ${^{\rm t}}$Reference [\onlinecite{Laude1971}].
\label{table3}
\end{table*}

For the $f$ process we consider the valleys at (0, 0, 0.8375) and (0, 0.8375, 0) \cite{Jacoboni2010} (in the [001] and [010] directions). Here they are the TO (green line) and secondary the LA (red line) modes that dominate, as indicated by the $f$-labelled point in Fig. \ref{fig6}(e). Note that in this case, the final states are the CBM states in the $\Gamma$-K direction with ~116\% of half Brillouin zone length away. The matrix elements near the $\Gamma$ point, indicating the intravalley scattering, have strong contributions from the acoustic phonon modes as also shown in Fig. \ref{fig5}(h), and the TA1 and LA are both contributing, with a weaker contribution from the TA2. The increase in the TA contributions in this $\Gamma$-K compared to the $\Gamma$-X direction, is a signature of the shear mode component that is now important. In Fig. \ref{fig6}f we also show the matrix elements in the $\Gamma$-L direction. In this case, the transitions only involve the intravalley scattering, where the only important ones are located around the $\Gamma$ point.

\begin{figure*}[tbp]
\includegraphics[width=0.7\textwidth]{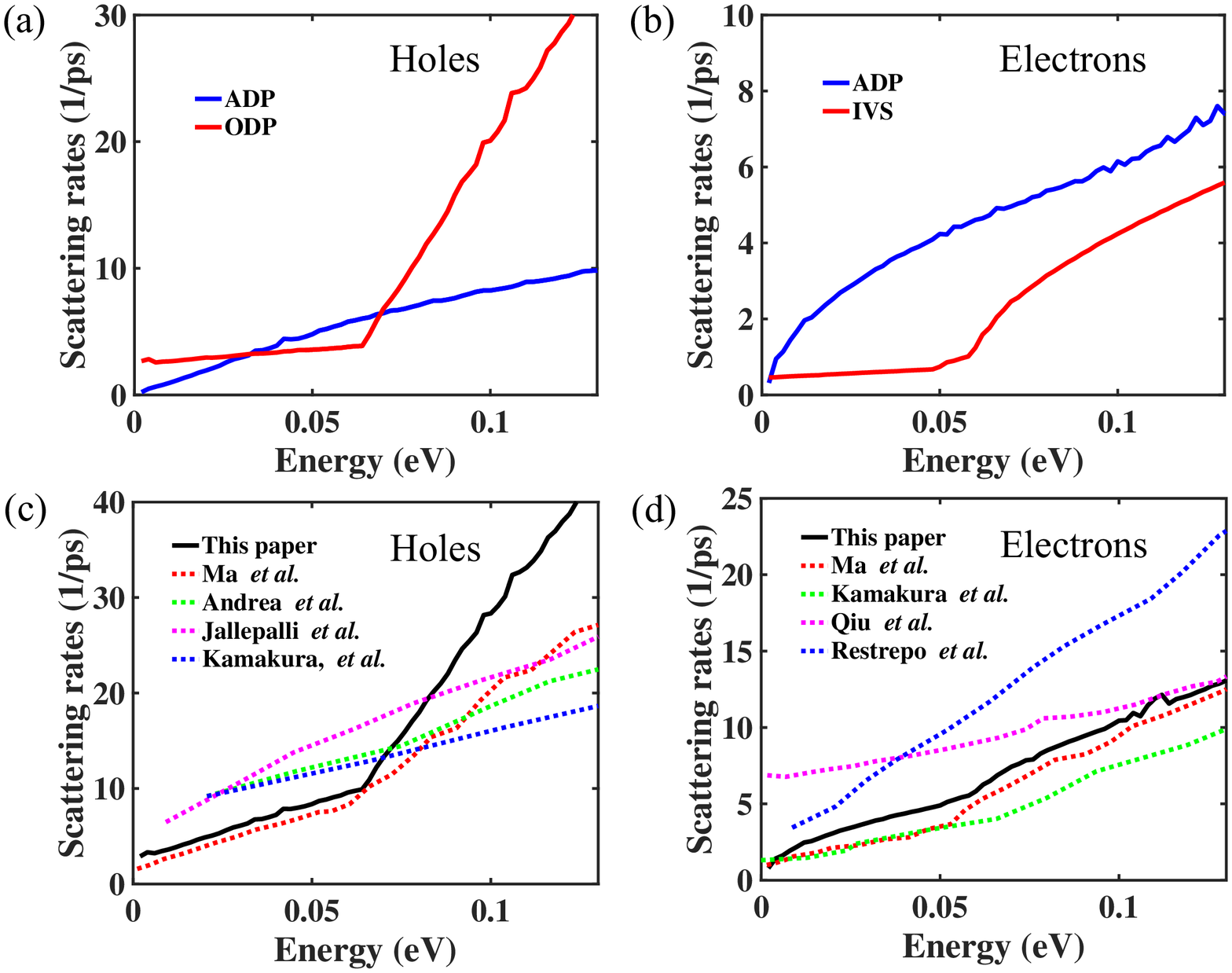}
\caption{Scattering rate of Si at 300 K vs energy for (a) holes under ADP and ODP scattering, (b) electrons under ADP and IVS scattering. The total scattering rates for (c) holes and (d) electrons in comparison with other $ab$ $initio$ results from Refs. \onlinecite{Ma2018,Qiu2015,Restrepo2009}  and empirical results from Refs. [\onlinecite{Ghetti2003,Jallepalli1997,Kamakura2000,Kamakura1994}].}
\label{fig7}
\end{figure*}

It is worth noting that also experimentally, the $g$-type scattering is identified to be caused by LO phonons, while the $f$-type scattering to be caused by LA and TO phonons \cite{Streitwolf1970}. Early reports mentioned that low-energy LA phonons can also be involved in $g$-type scattering \cite{Onto1969,Costato1970}. However, electron-phonon coupling selection rules \cite{Lax1961,Lax1972} do not allow $g$-type scattering by LA phonons, as can be verified by the matrix elements in Fig. \ref{fig6}d. This means that in the modeling of Si electrons transport, acoustic phonon scattering is exclusively considered for intravalley scattering, LO phonon scattering exclusively for intervalley $g$ processes, and TO and LA scattering for intervalley $f$ processes.

For $f$-type scattering, apart from scattering with the final states in the same Brillouin zone, we also consider scattering into the adjacent Brillouin zone (see Appendix F). It is found that the dominant modes keep the same for both the $g$-type and $f$-type scattering into the same or different Brillouin zones, where the values of deformation potentials of intervalley transitions are also comparable, even though the wave vectors are different. For the $f$-type scattering, the extracted deformation potential values of both the LA and TO modes, which are the dominant phonons, have less than 2\% difference for the scattering into the same or different Brillouin zone. For scattering into the same Brillouin zone, the intervalley deformation potentials for LA and TO modes are 1.86 and 3.59 eV/\AA, respectively, while the values for transitions into the second Brillouin zone are 1.83 and 3.55 eV/\AA, respectively.

\section{Transport properties}

Considering all the possible e-ph scattering processes, we compute the transport properties of Si. For holes, we consider ADP and ODP, while for electrons we consider ADP and IVS. The values of deformation potential used for ADP (eV), ODP (eV/\AA), IVS (eV/\AA) and phonon frequencies (meV) are listed in Table III. All the transport calculations are conducted using our own-developed Boltzmann transport equation simulator ElecTra \cite{Patrizio2021}, whose details can be found in the previous papers \cite{Graziosi2019,Neophytou2020,Graziosi2020,Graziosi20202}, which discretizes the 3D dispersion and constructs scattering times for every transport state using the derived deformation potentials.

To obtain an indication of the phonon-limited scattering rates, we combine different scattering processes for all bands with band index $i$ into one global rate (at room temperature) by averaging the rates with the band density of states as:
\begin{eqnarray}
|S_{\bf{k,k'}}|=\frac{|S_{\bf{k,k'}}^i| \textsl{g}_{\bf k'}^i}{\sum{\textsl{g}_{\bf k'}^i}},
\end{eqnarray}
where $\textsl{g}_{\bf k'}^i$ is the density of states for band $i$. Figure 7(a) shows the contributions of ADP and ODP to the scattering of holes versus energy. At low energies both ADP and ODP influence the rate, whereas at energies above 0.064 eV the ODP scattering rate greatly increases, as the phonon emission process is activated for energies above $\hbar \omega$. Figure 7(b) shows the contributions of ADP and IVS to the scattering of electrons. IVS in this case behaves like ODP since the phonon momentum required is large for the intervalley phonon energy to be assumed constant. We observe that the emission energy for IVS of electrons is lower than that for ODP in the case of holes. This is because near the zone boundary the energies of both acoustic and optical phonons that take part are comparable and are somewhat smaller than the longitudinal optical phonon energy at $\Gamma$. The calculated total scattering rates of holes and electrons are very similar to the EPW calculations by Ma $et al.$ \cite{Ma2018}, as seen in Figure 7(c) and 7(d), except for holes at higher energies that our results deviate somewhat, although they follow the same trend. Our scattering rates are also comparable to those of other $ab$ $initio$ calculations \cite{Qiu2015,Restrepo2009}  and empirical results as well \cite{Ghetti2003,Jallepalli1997,Kamakura2000,Kamakura1994}.

\begin{figure}[tbp]
\includegraphics[width=0.4\textwidth]{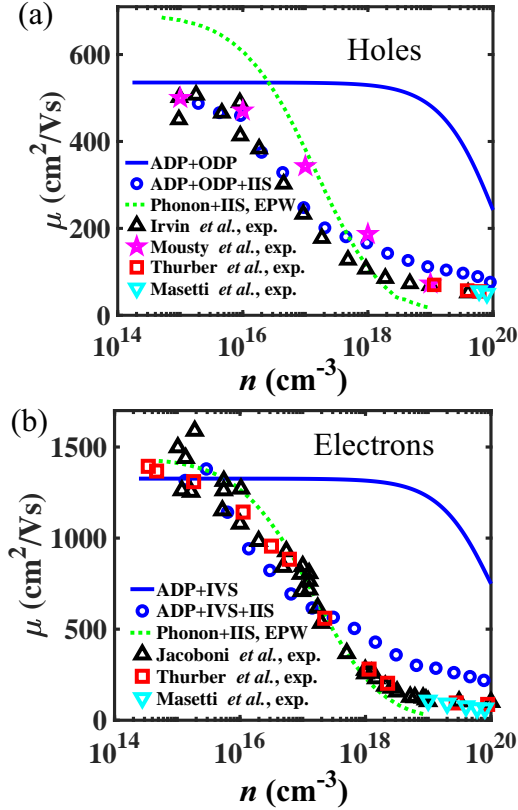}
\caption{(a) Mobility for holes in Si at 300 K calculated in this paper with solid (phonon-limited) and circled (phonon plus IIS) lines, compared to that using EPW for e-ph scattering from Ref. [\onlinecite{Ponce2018}], in which IIS is described using a phenomenological $ad$ $hoc$ equation in the effective mass approximation. Comparison is also made with experimental mobility values from Refs. [\onlinecite{Irvin1962,Mousty1974,Thurber1981,Masetti1983}]. (b) Mobility for electrons in Si at 300 K calculated in this paper with solid (phonon-limited) and circled (phonon plus IIS) lines, compared to that using EPW for e-ph scattering from Ref. [\onlinecite{Ponce2018}] and experiments from Refs [\onlinecite{Jacoboni1977,Thurber1981,Masetti1983}]. For the case of the blue circled lines (phonon plus IIS), we use the Brooks-Herring model to compute IIS. Notice that one blue circle for the case of phonon plus IIS at low carrier concentrations is slightly higher compared to the phonon-limited mobility due to the usual large amount of numerical noise associated with the IIS scattering rates \cite{Qiu2015}.}
\label{fig8}
\end{figure}

Figures \ref{fig8}a and \ref{fig8}b show the comparison between our calculated mobilities (solid blue and dotted lines) and the measured ones at 300 K, as well as that from $ab$ $initio$ calculations using EPW \cite{Ponce2018}, as a function of carrier concentration between $10^{14}$ and $10^{20}$ cm$^{-3}$. In this case, in addition to the $ab$ $initio$ e-ph scattering, the Brooks and Herring model \cite{Jacoboni1983} is used to describe the elastic scattering rate due to ionized dopants. We compare our calculations to multiple experiments over a range of doping levels. The scattering rate due to the ionized impurity scattering (IIS) is then given by
\begin{eqnarray}
|S_{\bf{k,k'}}^{\rm{IIS}}|=\frac{2\pi}{\hbar} \frac{Z^2 e^4}{\varepsilon_r^2 \varepsilon_0^2} \frac{N_{imp}}{(|{\bf k}- {\bf k'}|^2+\frac{1}{L_D^2})^2} \textsl{g}_{\bf{k'}},
\label{eq24}
\end{eqnarray}
where $Z$ is the electric charge of the ionized impurity, $\varepsilon_r$ and $\varepsilon_0$ are the relative and vacuum permittivities, $N_{imp}$ is the density of the ionized impurities, and $L_D$ is the screening length defined as
\begin{eqnarray}
L_D=\sqrt{\frac{\varepsilon_r \varepsilon_0}{e} (\frac{\partial E_F}{\partial n})},
\label{eq25}
\end{eqnarray}
where $E_F$ is the Fermi level and $n$ is the carrier density.

For holes, the phonon-limited mobility (blue solid line) at low densities (left sides of Fig. \ref{fig8}) is calculated to be 536 cm$^2$/V s. Despite this value being somewhat higher than measured \cite{Ludwig1956,Jacoboni1977,Cronemeyer1957}, it is known that $ab$ $initio$ calculations in general overestimate the Si hole mobility, and in fact the EPW results are significantly higher \cite{Ponce2018}. For electrons, our calculated phonon-limited mobility is 1327 cm$^2$/V s (blue solid line), in good agreement with previous works and measurements \cite{Ma2018,Li2015,Qiu2015,Ponce2018,Ludwig1956,Jacoboni1977,Li1977,Cronemeyer1957} (see Appendix G for comparison to other results). Overall, our calculated mobilities from the deformation potentials we derived agree well with that measured in experiment. A slight overestimation of our phonon plus IIS-limited results compared to measurements is observed at high carrier concentrations, where our calculated mobilities with IIS are somewhat larger than the measured ones for both holes and electrons. This is consistent with the previous works \cite{Fiorentini2016,Chattopadhyay1981} using the same Brooks-Herring model. On the other hand, the mobilities with IIS in the EPW work \cite{Ponce2018} are lower than the measured ones at high carrier concentrations because semi-empirical models are used to account for IIS to match the mobility trend. In general, at those carrier densities, it is possible and claimed that the electron-electron scattering, as well as additional dopant-specific considerations about IIS, which is not included in the Brooks-Herring model, provides an additional scattering mechanism to reduce the mobility even further to map experiments better \cite{Kosina1997,Kosina2018}. 

\section{Discussion}

The primary material focus of this paper was the well-established Si. However, the intent of this paper stretches far beyond Si, and it is to present a computationally efficient method (much more efficient compared to the fully first-principles calculations of the matrix elements) for mobility calculations, still with adequate first-principles accuracy. The fact that Si was the material of choice is the vast availability of deformation potential values and mobility data to benchmark our calculations on. The method developed in this paper can be applied to semiconductors with more complicated band structures, lower symmetry, and larger unit cells compared to Si, in which cases the computational savings can be quite significant. The numerical cost of using this method is much smaller than computing a huge number of matrix elements throughout the Brillouin zone (e.g., requiring 40$\times$40$\times$40 = 64 000 phonon $\bf q$ points \cite{Ponce2018}) as in common fully $ab$ $initio$ methods like EPW. Our method only needs a limited number of matrix elements (requiring a few $\bf q$ points, i.e., $\sim$100 only) around the VBM/CBM to derive deformation potentials, no matter the structure of the bands, whether that being a simple material with high symmetry and a few phonon modes, or a complicated material with a lot of optical phonon modes. For acoustic phonons in a more complex material the process still requires one LA and two TA branches as in Si (and all other semiconductor materials), although the numerical cost is slightly higher than Si due to the possibly larger number of initial/final bands. The number of optical matrix elements will increase if more optical phonon modes are present, but still, only a few matrix elements are needed per phonon mode (and in the case where the modes are not flat, a few more might be needed to provide an acceptable average), which makes it computationally feasible to carry on the calculation for all initial/final states and phonon branches for both intra/intervalley transitions, even without considering symmetries to reduce computation. Of course, it will be convenient in that case to devise an automated way to identify all VBM/CBM and phonon modes to avoid manual band structure exploration, and this is something we are currently investigating.

We note that our method is based on the deformation potential theory, which is proposed by Bardeen and Shockley \cite{Bardeen1950} for nonpolar semiconductors and insulators and recently extended to polar materials by excluding the Fröhlich interactions from the overall matrix elements \cite{Giustino2007}. In the case of calculating the electronic conductivity of metals, on the other hand, the usual method is using the Eliashberg function (or spectral distribution function of electron-phonon interaction), which is essentially the phonon density of states weighted by the electron-phonon coupling matrix element \cite{Allen1971,Savrasov1996,Hellsing2002}. As in the case of semiconductors, this is not an easy computation and it involves similar electron/phonon dispersions and calculations throughout the Brillouin zone, again the limiting factor being the number of $\bf q$ points included in the computation. However, the matrix elements can also be used to define deformation potentials under certain approximations, at least for the long-wavelength acoustic phonons, as described in the literature \cite{Khan1984,Kartheuser1986}. A formalism that exchanges the matrix elements in the spectral distribution function by effective deformation potentials could reduce the computation cost to the levels that we describe in this paper, as similarly it will not be necessary to compute all matrix elements throughout all the Brillouin zone.

One case where matrix elements in a larger part of the Brillouin zone could be needed, is high-field transport. Deformation potentials are used for high-field transport in common semiconductor devices for decades now with very good accuracy \cite{Fischetti1991,Fischetti19912}. In that case a more expanded set of matrix elements can be used to extract deformation potentials to increase accuracy. However, the computational cost will still be lower compared to computing matrix elements across the Brillouin zone.

Finally, we note that as in every DFT simulation, the outcome depends (sometimes sensitively) on the pseudopotentials and exchange correlation functionals. Prior works have quantified that for Si the intrinsic mobilities at 300 K differ by 16\% between LDA and GGA for electrons, but much less for holes by 3\% \cite{Ponce2018}. However, closer inspection showed that these differences arise primarily from the optimized lattice parameters obtained within these functionals, rather than the functionals themselves \cite{Ponce2018}. In particular, if the same lattice parameter is used in combination with different functionals, then the differences in the mobility are insignificant at 0.4\% for electrons and 2\% for holes \cite{Ponce2018}. To confirm this we have calculated and compared the matrix elements using the GGA-PBE-norm-conserving (the most commonly employed in EPW), GGA-PBE-PAW, GGA-PBEsol-PAW, and LDA-PZ-PAW pseudopotentials \cite{Andrea2014}. Using a common relaxed lattice parameter 5.479 \AA\ from the GGA-PBE-norm-conserving, the deviation of matrix elements between different pseudopotentials and exchange-correlation functionals is at most 4\% for LA and 1\% for LO phonons (see Table IV in Appendix B), leading to mobility variations of a few percentage units only. This is consistent with the claim \cite{Ponce2018} that the choice of exchange and correlation is not critical to the mobility as long as accurate lattice parameters are employed.

\section{Conclusions}

Based on density-functional theory (DFT) and density-functional perturbation theory (DFPT), we have developed a first-principles framework to extract acoustic, optical, and intervalley deformation potentials from the short-range electron-phonon (e-ph) matrix elements, for incorporation with the Boltzmann transport equation (BTE). Using the BTE based on a numerical simulator that allows for the incorporation of e-ph scattering and ionized impurity scattering (IIS), we are able to compute a comparable mobility with results from advanced first-principle calculations. The method we present would be the middle ground computationally between the constant relaxation time (CRT) approximation and $ab$ $initio$ relaxation time extraction with ultra-dense grids, while providing first principles accuracy. Although we have used Si as the material of investigation, the method can be generalized and applied to other solid-state semiconductors and insulators, with much higher computational efficiency compared to fully $ab$ $initio$ simulations.

\begin{acknowledgments}
This work has received funding from the European Research Council (ERC) under the European Union's Horizon 2020 research and innovation programme (Grant Agreement No. 678763). The authors thank for helpful discussions with Prof. William Vandenberghe from the University of Texas at Dallas. 
\end{acknowledgments}

\appendix

\section{Matrix elements from DFPT calculations}

From DFPT \cite{Baroni2001}, the displacement vector of the atoms with mass $m_k$, which are displaced from their equilibrium positions due to a phonon with mode $\nu$ and crystal momentum $\bf{q}$, is given by
\begin{eqnarray}
u_{lk\alpha}^{\nu {\bf q}} = \frac{1}{\sqrt{{m_k}}} e^{i{\bf q} \cdot {\bf R}_l} e_{\nu {\bf q}}^{k\alpha},
\label{eq26}
\end{eqnarray}
where $l$ labels the unit cell, $\alpha$ is the atom label, and $e_{\nu {\bf q}}^{k\alpha}$ are the phonon eigenvectors. At each position $\bf r$, the $\Delta_{\nu {\bf q}} V({\bf r})$, which is the perturbing potential due to phonon vibration, is calculated as
\begin{eqnarray}
\Delta_{\nu {\bf q}} V({\bf r}) = \sum_{lk\alpha} u_{lk\alpha}^{\nu {\bf q}}  \partial_{{\bf R}_l,k\alpha} V({\bf r}) \\ 
=  \sum_{k\alpha} \frac{1}{\sqrt{{m_k}}} e_{\nu {\bf q}}^{k\alpha} \partial_{{\bf q},k\alpha} V({\bf r}),
\label{eq27}
\end{eqnarray}
where $\partial_{{\bf q},k\alpha}V({\bf r}) $ is a term proportional to the derivatives of the Kohn-Sham potential $V({\bf r}) $ with respect to changes in the atomic positions ${\bf R}_{l,k\alpha}$ located at lattice vector ${\bf R}_l$ as
\begin{eqnarray}
\partial_{{\bf q},k\alpha}V({\bf r}) = \sum_{{\bf R}_l} e^{i{\bf q} \cdot {\bf R}_l} \partial_{{\bf R}_l,k\alpha} V({\bf r}),
\label{eq28}
\end{eqnarray}
where
\begin{eqnarray}
\partial_{{\bf R}_l,k\alpha} V({\bf r}) = \frac{\partial V({\bf r})} {\partial u_{lk\alpha}}.
\label{eq29}
\end{eqnarray}

\section{Choice of pseudopotentials and exchange-correlation functionals}

Taking $\bf q$ = (0.1, 0.1, 0.1) as an example, Table \ref{table4} lists the $M_{mn}^\nu ({\bf{k},\bf{q}})$ for the HH-HH transition of the VBM for the LA and LO phonon modes using different pseudopotentials and exchange-correlation functionals. The same lattice parameter 5.479 \AA\ relaxed from the GGA-PBE-norm-conserving pseudopotential is used. The difference of $M_{mn}^\nu ({\bf{k},\bf{q}})$ using different pseudopotentials is at most 4\% for the LA and 1\% for the LO modes. 

\begin{table}
\caption{$M_{mn}^\nu ({\bf{k},\bf{q}})$ (in eV/\AA) for the HH-HH transition of the VBM with $\bf q$ = (0.1, 0.1, 0.1) for the LA and LO phonon modes using different pseudopotentials and exchange-correlation functionals.}
\begin{ruledtabular}
\begin{tabular}{lcc}
Pseudopotentials & LA & LO \\
\hline
GGA-PBE-norm-conserving & 0.533 & 3.723\\
GGA-PBE-PAW & 0.529 & 3.727 \\
GGA-PBEsol-PAW & 0.527 & 3.696 \\
LDA-PZ-PAW & 0.512 & 3.697 \\
\end{tabular}
\end{ruledtabular}
\label{table4}
\end{table}

\section{Coupling matrix of transverse modes for holes}

\begin{figure}[tbp]
\includegraphics[width=0.35\textwidth]{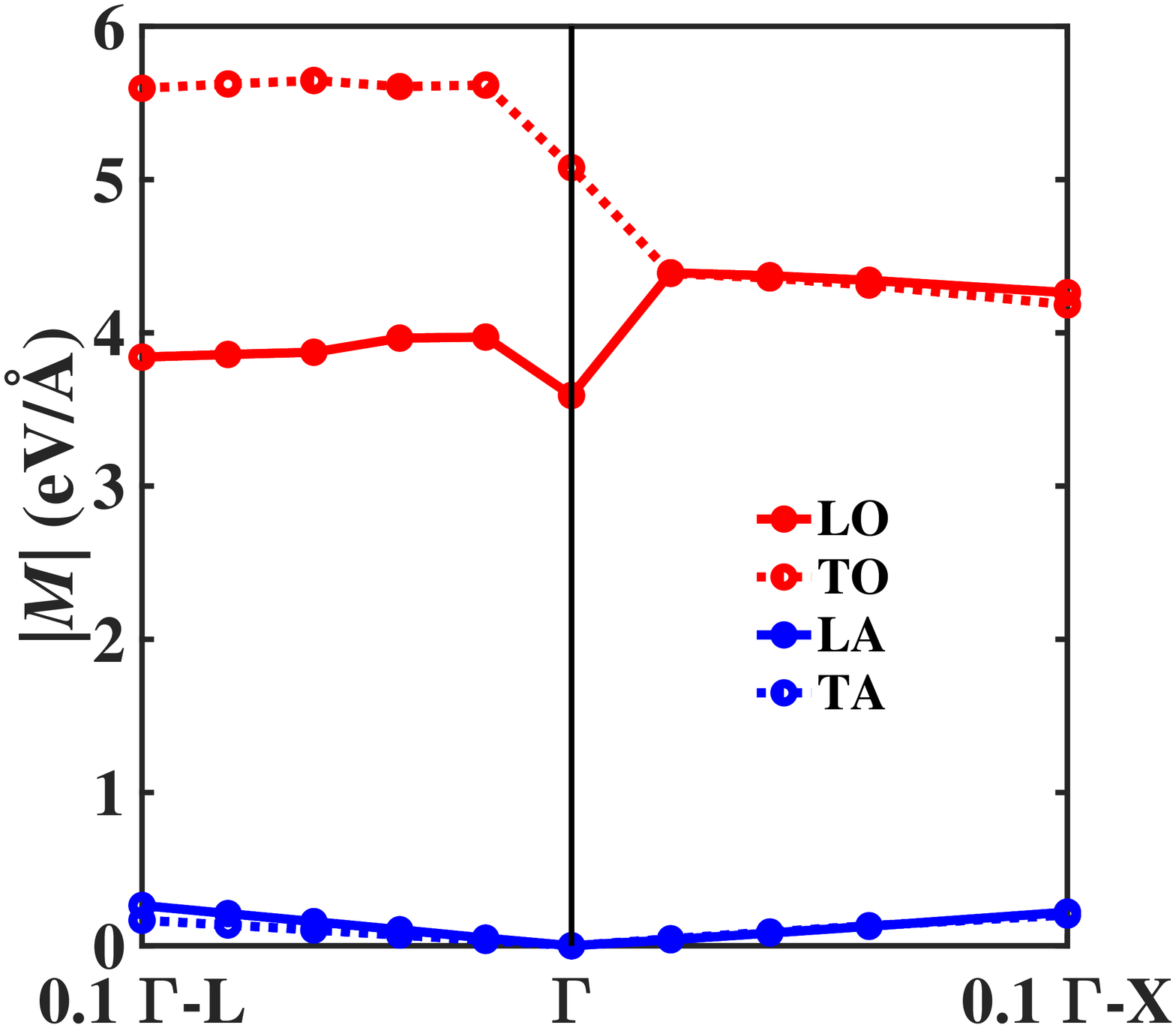}
\caption{The coupling matrix elements for LO, TO, LA, and TA modes near the $\Gamma$ point of holes for scattering into the HH of Si.}
\label{fig9}
\end{figure}

Figure \ref{fig9} shows the coupling matrix elements  $M_{mn}^\nu ({\bf{k},\bf{q}})$ for longitudinal and transverse modes of holes for scattering into the HH in Si near the $\Gamma$ point. Here the transverse modes consider both branches.

\section{Deformation potentials using the largest singular value}

In the finite-dimensional case, a matrix $M$ can be decomposed in the form $U\Sigma V^{\rm T}$, where $U$ and $V^{\rm T}$ are unitary matrices and $\Sigma$ is a diagonal matrix with the singular values residing on the diagonal. This procedure is called singular value decomposition (SVD). The diagonal entries $\sigma_i = S_{ii}$ are known as the singular values of $M$. The SVD is not unique and it is always possible to choose the decomposition so that the singular values $\sigma_i$ are in descending/ascending order. In this case, we can identify the largest singular value $\sigma_1$. For example, for the VBM of Si, there are three degenerate bands. Thus, we can use a $3 \times 3$ matrix for the nine deformation potential components that couple the three initial and three final bands. We define the initial and final bases as $\langle a|$, $\langle b|$, $\langle c|$, $|d\rangle$, $|e\rangle$, $|f\rangle$, respectively, and thus the deformation potentials matrix is
\begin{eqnarray}
M = \begin{bmatrix}
D_{aa} & D_{ab} & D_{ac}\\
D_{ba} & D_{bb} & D_{bc}\\
D_{ca} & D_{cb} & D_{cc}\\
\end{bmatrix}.
\label{eq35}
\end{eqnarray}
After performing the singular value decomposition, we find the $3 \times 3$ matrices $U$, $S$, and $V$:
\begin{eqnarray}
U = \begin{bmatrix}
U_{11} & U_{12} & U_{13}\\
U_{21} & U_{22} & U_{23}\\
U_{31} & U_{32} & U_{33}\\
\end{bmatrix},
\label{eq36}
\end{eqnarray}

\begin{eqnarray}
\Sigma = \begin{bmatrix}
S_{11} & S_{12} & S_{13}\\
S_{21} & S_{22} & S_{23}\\
S_{31} & S_{32} & S_{33}\\
\end{bmatrix},
\label{eq37}
\end{eqnarray}

\begin{eqnarray}
V = \begin{bmatrix}
V_{11} & V_{12} & V_{13}\\
V_{21} & V_{22} & V_{23}\\
V_{31} & V_{32} & V_{33}\\
\end{bmatrix}.
\label{eq38}
\end{eqnarray}
The singular values are the overall deformation potentials of the degenerate bands. In this case the overall deformation potential can be defined as the largest singular value $\sigma_1 = S_{11}$ (the others turn out to be zero). The original and new initial and final bases obey the following relations (where the primed vectors refer to the new basis):
\begin{eqnarray}
\begin{aligned}
& \langle a| =  V_{11} \langle a'| + V_{12} \langle b'| + V_{13} \langle c'|, \\
& \langle b| =  V_{21} \langle a'| + V_{22} \langle b'| + V_{23} \langle c'|, \\
& \langle c| =  V_{31} \langle a'| + V_{32} \langle b'| + V_{33} \langle c'|, \\
& |d\rangle = U_{11} |d'\rangle  + U_{12} |e'\rangle  + U_{13} |f'\rangle, \\
& |e\rangle =  U_{21} |d'\rangle  + U_{22} |e'\rangle  + U_{23} |f'\rangle,\\
& |f\rangle =  U_{31} |d'\rangle  + U_{32} |e'\rangle  + U_{33} |f'\rangle.
\label{eq39}
\end{aligned}
\end{eqnarray}

Considering the ADP from $\Gamma \rightarrow \Gamma + {\bf q}$ process as an example for hole scattering, and using the corresponding values from Table I, we can define
\begin{eqnarray}
M = \begin{bmatrix}
0.5270 & 2.3398 & 2.3398\\
0.5270 & 2.3398 & 2.3398\\
0.5270 & 2.3398 & 2.3398\\
\end{bmatrix}
\label{eq40}
\end{eqnarray}
After performing the singular value decomposition (for example, using the svd command in Matlab), We find $U$, $S$, and $V$ to be
\begin{eqnarray}
U = \begin{bmatrix}
-0.5774 & 0.8165 & 0\\
-0.5774 & -0.4082 & -0.7071\\
-0.5774 & -0.4082 & 0.7071\\
\end{bmatrix},
\label{eq41}
\end{eqnarray}
\begin{eqnarray}
\Sigma = \begin{bmatrix}
5.8035 & 0 & 0\\
0 & 0 & 0\\
0 & 0 & 0\\
\end{bmatrix},
\label{eq42}
\end{eqnarray}
\begin{eqnarray}
V = \begin{bmatrix}
-0.1573 & 0.9876 & 0\\
-0.6983 & -0.1112 & -0.7071\\
-0.6983 & -0.1112 & 0.7071\\
\end{bmatrix}.
\label{eq43}
\end{eqnarray}
The largest singular value is 5.8035, which will be used as the overall deformation potential for acoustic phonon transitions in the valence band. The original and new initial and final bases then become
\begin{eqnarray}
\begin{aligned}
& \langle a| = -0.1573 \langle a'| + 0.9876 \langle b'|,\\
& \langle b| = -0.6983  \langle a'| -0.1112 \langle b'| -0.7071 \langle c'|,\\
& \langle c| = -0.6983  \langle a'|  -0.1112 \langle b'| + 0.7071\langle c'|,\\
& |d\rangle = -0.5774 |d'\rangle  + 0.8165 |e'\rangle,\\
& |e\rangle = -0.5774 |d'\rangle  -0.4082 |e'\rangle  -0.7071 |f'\rangle,\\
& |f\rangle = -0.5774 |d'\rangle -0.4082 |e'\rangle  + 0.7071 |f'\rangle.
\label{eq44}
\end{aligned}
\end{eqnarray}

On the other hand, for the $\Gamma - {\bf q}/2  \rightarrow \Gamma + {\bf q}/2$ process, using the corresponding values from TABLE I, for LA we can define
\begin{eqnarray}
M = \begin{bmatrix}
1.8482 & 0 & 0\\
0 & 2.5825 & 2.5825\\
0 & 2.5825 & 2.5825\\
\end{bmatrix},
\label{eq45}
\end{eqnarray}
We find $U$, $S$, and $V$ to be
\begin{eqnarray}
U = \begin{bmatrix}
0 & 1 & 0\\
-0.7071 & 0 & -0.7071\\
-0.7071 & 0 & 0.7071\\
\end{bmatrix},
\label{eq46}
\end{eqnarray}
\begin{eqnarray}
\Sigma = \begin{bmatrix}
5.165 & 0 & 0\\
0 & 1.8482 & 0\\
0 & 0 & 0\\
\end{bmatrix},
\label{eq47}
\end{eqnarray}
\begin{eqnarray}
V = \begin{bmatrix}
0 & 1 & 0\\
-0.7071 & 0 & -0.7071\\
-0.7071 & 0 & 0.7071\\
\end{bmatrix}.
\label{eq48}
\end{eqnarray}
The original and new initial and final bases then become
\begin{eqnarray}
\begin{aligned}
& \langle a| = \langle b'|, \\
& \langle b| = -0.7071  \langle a'| -0.7071 \langle c'|, \\
& \langle c| = -0.7071  \langle a'| + 0.7071\langle c'|, \\
& |d\rangle = |e'\rangle, \\
& |e\rangle = -0.7071 |d'\rangle -0.7071 |f'\rangle, \\
& |f\rangle = -0.7071 |d'\rangle + 0.7071 |f'\rangle.
\label{eq49}
\end{aligned}
\end{eqnarray}
From $S$, we find two singular values, 5.165 and 1.8482. The largest singular value from $\langle a'|$ and $|d'\rangle$ is related to $\langle b|$, $\langle c|$, $|e\rangle$, and $|f\rangle$, which are related to the HH. The second-largest singular value is associated with the $\langle b'|$ and $|e'\rangle$, which are the same as the $\langle a|$ and $|d\rangle$, and are related to the transition process from LH to LH. Compared to the $\Gamma \rightarrow \Gamma + {\bf q}$, the $\Gamma - {\bf q}/2  \rightarrow \Gamma + {\bf q}/2$ has different values of coupling matrix elements for the transitions of different processes, especially for the values between the HH and LH transitions. These are finite and zero for the two processes, respectively. Using the largest singular value for the overall process, however, interestingly the overall deformation potentials are similar for both $\Gamma \rightarrow \Gamma + {\bf q}$ and $\Gamma - {\bf q}/2  \rightarrow \Gamma + {\bf q}/2$ processes. It is also interesting to observe that the second-largest value remains unchanged after singular value decomposition and it is equal to the LH-LH matrix element for the $\Gamma - {\bf q}/2 \rightarrow \Gamma + {\bf q}/2$ process [see Eqs. (\ref{eq45}) and (\ref{eq47})]. As observed from Eq. (\ref{eq45}) the two subspaces of LH and HH are independent and no transitions are allowed between them (zero off-diagonal elements connecting them).      

\section{Deformation potential with SOC}
The VBM of Si has three degenerate bands, when spin-orbit coupling (SOC) is omitted. If we label them as 1, 2, and 3, then we can find nine coupling matrix elements for all intertransitions $M_{ij}$ and form a global matrix as
\begin{eqnarray}
M = \begin{bmatrix}
M_{11} & M_{12} & M_{13}\\
M_{21} & M_{22} & M_{23}\\
M_{31} & M_{32} & M_{33}\\
\end{bmatrix}.
\label{eq30}
\end{eqnarray}

Taking a phonon wave vector $\bf q$ = (0.1, 0, 0) as an example, we can compute $M$ (without SOC) for LA and LO modes as
\begin{eqnarray}
{\rm LA} = \begin{bmatrix}
0.105 & 0.22 & 0.22 \\
0.105 & 0.22 & 0.22 \\
0.105 & 0.22 & 0.22 \\
\end{bmatrix},
\label{eq31}
\end{eqnarray}

\begin{eqnarray}
{\rm LO} = \begin{bmatrix}
0 & 4.26 & 4.26 \\
0 & 4.26 & 4.26 \\
0 & 4.26 & 4.26 \\
\end{bmatrix}.
\label{eq32}
\end{eqnarray}

In the case where we consider SOC, the coupling matrices $M_{ij}$ are computed as $\sqrt{(M_{ij})^2_{\uparrow\uparrow}+(M_{ij})^2_{\downarrow\downarrow}}$ and are found to be:

\begin{eqnarray}
{\rm LA} = \begin{bmatrix}
0.05 & 0.325 & 0 \\
0.13 & 0.133 & 0.268 \\
0.13 & 0.133 & 0.268 \\
\end{bmatrix},
\label{eq33}
\end{eqnarray}

\begin{eqnarray}
{\rm LO} = \begin{bmatrix}
0 & 0 & 6.02 \\
0.928 & 5.13 & 3 \\
0.928 & 5.13 & 3 \\
\end{bmatrix}.
\label{eq34}
\end{eqnarray}

Using singular value decomposition to choose the linear combinations of the spacial wave functions, we find the largest singular values to be 0.569 eV/\AA\ and 10.435 eV/\AA\ for LA and LO without SOC, respectively. With SOC, we find two singular values for each of the LA and LO, and the square root of the sum of the squares are 0.567 eV/\AA\ and 10.420 eV/\AA\ for LA and LO, respectively. Since the values with and without SOC are very similar, for simplicity, we use the three generate bands at VBM without SOC to derive the deformation potentials.

\section{Intervalley scattering for the conduction band}

\begin{figure}[tbp]
\includegraphics[width=0.35\textwidth]{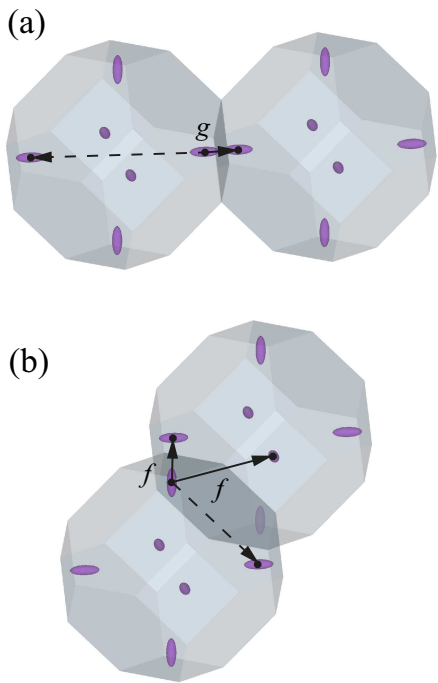}
\caption{The (a) $g$ and (b) $f$ processes shown in 3D Brillouin zones to identify the wave vectors of the phonons involved. The wave vectors to the same and different Brillouin zones are illustrated with the dashed and solid lines, respectively.}
\label{fig10}
\end{figure}

From geometrical considerations, the $g$ and $f$ scattering can happen between adjacent Brillouin zones \cite{Long1960,Sinha2005}. There are in total one $g$-type and four $f$-type scattering transitions, considering all the neighboring CBM valleys in the different Brillouin zones and symmetry restrictions. The wave vectors involved in these two types of electron transitions are illustrated in Fig. \ref{fig10}, where the dashed and solid lines show transitions with the final states residing in the same and different Brillouin zones, respectively. Here we set the vector for $f$ scattering as (0.1625, 0.1625, 1), which is nearly 13 degrees off the [001] direction.

It is found that the LO mode for $g$-type scattering, and LA and TO modes for $f$-type scattering, are always the dominant phonon modes no matter if we choose the wave vectors for the transitions towards the same or different Brillouin zones. The values of deformation potentials of intervalley transitions in the same or different Brillouin zones are also comparable, even though the wave vectors are different. For example, for $g$-type scattering within the same Brillouin zone, the intervalley deformation potentials for LO mode are 3.87 eV/\AA, while the values for transitions into the second Brillouin zone are 3.86 eV/\AA.

\section{Phonon-limited mobility}

Table \ref{table5} lists the phonon-limited mobility for holes and electrons of Si, where our calculated mobility is compared with other first-principles calculations \cite{Ma2018,Ponce2018,Li2015,Qiu2015} and experiments \cite{Ludwig1956,Jacoboni1977,Cronemeyer1957,Li1977}.

\begin{table}
\caption{The phonon-limited mobility (cm$^2$/V s) of holes and electrons for Si at 300 K in comparison to previous $ab$ $initio$ calculations and experiments.}
\begin{ruledtabular}
\begin{tabular}{lcc}
 & Holes & Mobility\\
\hline
This paper & calc. & 536 \\
Ma $et al.$ \cite{Ma2018} & calc. & 569 \\
P$\rm \acute{o}$nce $et al.$ \cite{Ponce2018} & calc. & 658 \\
Ludwig $et al.$ \cite{Ludwig1956} & exp. & 480 \\
Jacoboni $et al.$ \cite{Jacoboni1977} & exp. & 450 \\
Cronemeyer $et al.$ \cite{Cronemeyer1957} & exp. & 510 \\
\hline
 & Electrons & Mobility\\
\hline
This paper & calc. & 1327 \\
Ma $et al.$ \cite{Ma2018} & calc. & 1915 \\
Li $et al.$ \cite{Li2015} & calc. & 1860 \\
Qiu $et al.$ \cite{Qiu2015} & calc. & 1500 \\
P$\rm \acute{o}$nce $et al.$ \cite{Ponce2018} & calc. & 1366 \\
Ludwig $et al.$ \cite{Ludwig1956} & exp. & 1350 \\
Jacoboni $et al.$ \cite{Jacoboni1977} & exp. & 1450 \\
Li $et al.$ \cite{Li1977} & exp. & 1430 \\
Cronemeyer $et al.$ \cite{Cronemeyer1957} & exp. & 1360 \\
\end{tabular}
\end{ruledtabular}
\label{table5}
\end{table}

\bibliography{apssamp}

\end{document}